\documentclass[a4paper,11pt]{article}
\usepackage{amssymb}
\usepackage{amsmath}
\usepackage{latexsym,amssymb,amsmath}
\usepackage{graphicx}
\usepackage{epsfig}

\DeclareFontFamily{U}{rsf}{} \DeclareFontShape{U}{rsf}{m}{n}{
  <5> <6> rsfs5 <7> <8> <9> rsfs7 <10-> rsfs10}{}
\DeclareMathAlphabet\Scr{U}{rsf}{m}{n} \makeatletter
\@addtoreset{equation}{section} \makeatother

\newcommand{\be}{\begin{equation}}
\newcommand{\ee}{\end{equation}}
\newcommand{\bea}{\begin{eqnarray}}
\newcommand{\eea}{\end{eqnarray}}
\newcommand{\ba}{\begin{array}}
\newcommand{\ea}{\end{array}}
\newcommand{\bit}{\begin{itemize}}
\newcommand{\eit}{\end{itemize}}
\newcommand{\ben}{\begin{enumerate}}
\newcommand{\een}{\end{enumerate}}

\begin{document}

\begin{titlepage}
 \thispagestyle{empty}
\begin{flushright}
     \hfill{CERN-PH-TH/2010-224}\\
     \hfill{SU-ITP-10/31}
 \end{flushright}

 \vspace{50pt}

 \begin{center}
     { \huge{\bf      {Split Attractor Flow \\  \vspace{5pt}in $\mathcal{N}=2$ Minimally Coupled Supergravity
    }}}

     \vspace{50pt}

     {\Large {Sergio Ferrara$^{a,b,c}$, Alessio Marrani$^{d}$ and Emanuele Orazi$^{a}$ }}

     \vspace{30pt}

  {\it ${}^a$ Physics Department, Theory Unit, CERN,\\
     CH -1211, Geneva 23, Switzerland;\\
     \texttt{sergio.ferrara@cern.ch}\\
     \texttt{emanuele.orazi@polito.it}}

     \vspace{10pt}

    {\it ${}^b$ INFN - Laboratori Nazionali di Frascati,\\
     Via Enrico Fermi 40, I-00044 Frascati, Italy}

     \vspace{10pt}

     {\it ${}^c$  Department of Physics and Astronomy,\\
University of California, Los Angeles, CA 90095-1547,USA}\\

     \vspace{10pt}

   {\it ${}^d$ Stanford Institute for Theoretical Physics,\\
     Stanford University, Stanford, CA 94305-4060,USA;\\
     \texttt{marrani@lnf.infn.it}}

     %\vspace{15pt}

     \vspace{50pt}

     {ABSTRACT}

 \vspace{10pt}
 \end{center}
We classify the stability region, marginal stability walls (MS) and split attractor flows for two-center extremal black
holes in four-dimensional $\mathcal{N}=2$ supergravity \textit{minimally coupled} to $%
n$ vector multiplets.

It is found that two-center (continuous) charge orbits, classified
by four duality invariants, either support a stability region ending
on a MS wall or on an anti-marginal stability (AMS) wall, but not
both. Therefore, the scalar manifold never contains both walls.
Moreover, the BPS mass of the black hole composite (in its stability
region) never vanishes in the scalar manifold. For these reasons,
the ``bound state transformation walls'' phenomenon does not
necessarily occur in these theories.

The entropy of the flow trees also satisfies an inequality which
forbids ``entropy enigma'' decays in these models.

Finally, the non-BPS case, due to the existence of a ``fake''
superpotential satisfying a triangle inequality, can be treated as
well, and it can be shown to exhibit a split attractor flow dynamics
which, at least in the $n=1$ case, is analogous to the BPS one.
\end{titlepage}
\tableofcontents
\section{\label{Intro}Introduction}

The present paper is devoted to the study of the two-center extremal black
hole (BH) solution and split attractor flow \cite{D-1} in $\mathcal{N}=2$, $%
d=4$ supergravity \textit{minimally coupled} to $n$ Abelian vector
multiplets \cite{Luciani}. Within such a theory, the entropy of a
single-center extremal BH with dyonic charge vector\footnote{%
Note that the ``physically sensible'' charges are actually given by $%
\mathcal{Q}/\sqrt{2}$ and $\mathcal{X}/\sqrt{2}$ where $\mathcal{X}$ and $%
\mathcal{Q}$ are real and complex parameterizations of the charges,
respectively defined in (\ref{Q-def}) and (\ref{X-def}) further below.} $%
(p^{0},p^{i},q_{0},q_{i})$ is given by
\begin{equation}
\frac{S}{\pi }=\frac{1}{2}\left| \mathcal{I}_{2}\left( \mathcal{Q}\right)
\right| =\frac{1}{2}\left| p_{0}^{2}+q_{0}^{2}-p_{i}^{2}-q_{i}^{2}\right| \,,
\label{first}
\end{equation}
with $\mathcal{I}_{2}\gtrless 0$ for BPS and non-BPS solutions,
respectively. Note that Eq. (\ref{first}) reduces to the
Reissner-Nordstr\"{o}m BPS BH entropy if one sets $p^{i}=q_{i}=0$.
However, the ADM mass \cite{ADM} depends on scalars, due to
the presence of the $e^{\mathcal{K}/2}$ K\"{a}hler factor in the $\mathcal{N}%
=2$ central charge function:
\begin{equation}
Z\left( \left| t^{i}\right| ^{2};p_{0},q_{0}\right) =\frac{\left(
q_{0}+i\,p_{0}\right) }{\sqrt{2\left( 1-\left| t^{i}\right| ^{2}\right) }}\,.
\end{equation}

Two-center solutions exist as well, with general different properties with
respect to the single-center cases. As we will show in the subsequent
treatment, a peculiar feature of the $\mathcal{N}=2$ \textit{minimally
coupled} is that AMS walls, when they exist, are not supported by charge
configurations which admit a split attractor flow. Moreover, within such
configurations, the single-center entropy with charge $\mathcal{Q}_{1}+%
\mathcal{Q}_{2}$ is always larger than the corresponding two-center entropy,
namely:
\begin{equation}
S\left( \mathcal{Q}_{1}+\mathcal{Q}_{2}\right) >S\left( \mathcal{Q}%
_{1}\right) +S\left( \mathcal{Q}_{2}\right) \,.  \label{EntropyBound}
\end{equation}
The inequality (\ref{EntropyBound}) implies that the ADM masses of the
constituents, as well as the one of the composite solution, are always
bounded from below in the scalar manifold. As a consequence, ``entropy
enigma decays'' \cite{BD-1,DM-1,DM-2,David} do not occur, and the bound
states do not necessarily have ``recombination walls'' \cite{ADMJ-2}.

The non-BPS branch can be investigated as well, exhibiting a split dynamics
analogous to the BPS case. However, an important difference with respect to
the BPS case is the presence of a \textit{``moduli space''} of non-BPS
solutions \cite{Ferrara-Marrani-2}. This is ultimately due to the fact that
non-BPS attractor equations (given by (\ref{pa2}) further below) define
hyper-planes, and not points \cite{Ferrara-Marrani-2,Gnecchi-1}.\medskip

The plan of the paper is as follows.

Sec. \ref{Basics} presents some basic facts on the geometric structure and
on the duality symmetries of $\mathcal{N}=2$, $d=4$ supergravity \textit{%
minimally coupled} to $n$ vector multiplets, which will then be exploited in
the subsequent treatment of split flow in this theory.

In Sec. \ref{n=1} we analyse the one-modulus case, namely the model which is
electric-magnetic dual to the \textit{axion-dilaton} model obtained as a
truncation (to two different $U\left( 1\right) $'s) of ``pure'' $\mathcal{N}%
=4$, $d=4$ supergravity (for a review and a list of Refs., see \textit{e.g.}
\cite{FHM-rev}). The corresponding non-BPS branch is studied in Sec. \ref
{Axion-Dilaton-nBPS}.

Sec. \ref{n>1} extends the analysis of the BPS two-center split flow to an
arbitrary number of Abelian vector multiplets.

In Sec. \ref{BPS-t^3-Analysis} a comparison with the so-called $\mathcal{N}%
=2 $, $d=4$ $t^{3}$ model is worked out.

The paper ends with some comments and remarks in\ Sec. \ref{Conclusion},
along with a couple of Appendices, providing some technical details on the
MS and AMS conditions for the split scalar flows.

\section{\label{Basics}Basics}

The scalar manifold of the $\mathcal{N}=2$, $d=4$ \textit{minimally coupled}
Maxwell-Einstein supergravity theory provides the simplest example of
symmetric special K\"{a}hler space, which is locally a (non-compact version
of the) $\mathbb{CP}^{n}$ space:
\begin{equation}
\frac{SU\left( 1,n\right) }{SU\left( n\right) \times U\left( 1\right) }.
\end{equation}
The main feature of the corresponding special geometry is the vanishing of
the tensor $C_{ijk}$, yielding the following Riemann and Ricci tensors (see
\textit{e.g.} \cite{CDF-rev}, and Refs. therein)
\begin{equation}
R_{i\overline{j}k\overline{l}}=-g_{i\overline{j}}g_{k\overline{l}}-g_{i%
\overline{l}}g_{k\overline{j}}\Rightarrow R_{i\overline{j}}=-\left(
n+1\right) g_{i\overline{j}}.
\end{equation}

The special coordinates preserving the $SU\left( 1,n\right) $ symmetry are
based on the holomorphic prepotential function
\begin{equation}
F\left( X\right) =-\frac{i}{2}\left[ \left( X^{0}\right) ^{2}-\left(
X^{i}\right) ^{2}\right] \equiv \left( X^{0}\right) ^{2}\mathcal{F}\left(
t\right) ,  \label{prim-1}
\end{equation}
such that the holomorphic symplectic sections read ($F_{\Lambda }\left(
X\right) \equiv \frac{\partial F}{\partial X^{\Lambda }}$, $\Lambda
=0,1,...,n$ throughout)
\begin{eqnarray}
\mathbf{V} &=&\left( X^{\Lambda },F_{\Lambda }\left( X\right) \right)
^{T}=\left( X^{0},X^{i},-iX^{0},iX^{i}\right) ^{T}=  \notag \\
&=&\left( 1,t^{i},-i,it^{i}\right) ,  \label{prim-2}
\end{eqnarray}
where in (\ref{prim-1}) and in the second line of (\ref{prim-2}) projective
coordinates $t^{i}\equiv X^{i}/X^{0}$ have been introduced, with $%
X^{0}\equiv 1$ eventually fixed by choosing a suitable K\"{a}hler gauge.
Correspondingly, the covariantly holomorphic symplectic sections read
\begin{equation}
\mathcal{V}\equiv \left( L^{\Lambda },M_{\Lambda }\right) ^{T}\equiv e^{%
\mathcal{K}/2}\mathbf{V},  \label{V-call}
\end{equation}
where the K\"{a}hler potential $\mathcal{K}$ is then given by (see \textit{%
e.g.} \cite{CDF-rev}, and Refs. therein)
\begin{equation}
\mathcal{K}=-\ln \left[ i\left( \overline{X}^{\Lambda }F_{\Lambda
}-X^{\Lambda }\overline{F}_{\Lambda }\right) \right] =-\ln \left[ 2\left(
1-\left| t^{i}\right| ^{2}\right) \right] .  \label{K-n}
\end{equation}
Note that, as a consequence of $C_{ijk}=0$, the special geometry relations
are very simple:
\begin{equation}
\overline{D}_{\overline{i}}\mathcal{V}=0,~D_{i}D_{j}\mathcal{V}=0,~\overline{%
D}_{\overline{j}}D_{i}\mathcal{V}=g_{i\overline{j}}\mathcal{V},
\end{equation}
where $D_{i}$ and $\overline{D}_{\overline{i}}$ respectively denote the
K\"{a}hler-covariant differential operators, whose action on $\mathbf{V}$
reads
\begin{equation}
D_{i}\mathbf{V}=\left( \partial _{i}+\partial _{i}\mathcal{K}\right) \mathbf{%
V},~~\overline{D}_{\overline{i}}\mathbf{V}=\overline{\partial }_{\overline{i}%
}\mathbf{V}=0.
\end{equation}

The scalar-dependent central extension $Z$ (\textit{central charge}) of the $%
\mathcal{N}=2$ local supersymmetry algebra is built from the symplectic
product of the dyonic vector of (magnetic $p$ and electric $q$) charges of
the two-form field strengths
\begin{equation}
\mathcal{Q}\equiv \left( p^{0},p^{i},q_{0},q_{i}\right) ^{T}  \label{Q-def}
\end{equation}
and of the vector of covariantly holomorphic symplectic sections $\mathcal{V}
$ (\ref{V-call}) as follows:
\begin{eqnarray}
Z &\equiv &\left\langle \mathcal{Q},\mathcal{V}\right\rangle =\mathcal{Q}%
^{T}\Omega \mathcal{V}=q_{0}L^{0}+q_{i}L^{i}-p^{0}M_{0}-p^{i}M_{i}=e^{%
\mathcal{K}/2}\left( q_{\Lambda }X^{\Lambda }-p^{\Lambda }F_{\Lambda }\right)
\notag \\
&=&\frac{\left[ q_{0}+ip^{0}+\left( q_{i}-ip^{i}\right) t^{i}\right] }{\sqrt{%
2}\sqrt{1-\left| t^{k}\right| ^{2}}}.  \label{Z-CP^n}
\end{eqnarray}
where $\Omega $ is the $Sp\left( 2n+2,\mathbb{R}\right) $-metric. The
corresponding K\"{a}hler covariant derivatives (also named \textit{matter
charges}) read as follows \label{Gnecchi-1}:
\begin{equation}
D_{i}Z\equiv Z_{i}=\frac{\left[ (q_{i}-ip^{i})(1-\left| t^{l}\right|
^{2})+(q_{0}+ip^{0})\overline{t}^{\overline{i}}+(q_{j}-ip^{j})t^{j}\overline{%
t}^{\overline{i}}\right] }{\sqrt{2}\left( 1-\left| t^{k}\right| ^{2}\right)
^{3/2}}.  \label{DiZ-CP^n}
\end{equation}
It is also convenient to switch to a complex parametrization of the charge
vector (in the fundamental irrepr. $\mathbf{1+n}$ of $U\left( 1,n\right) $):
\begin{equation}
\mathcal{X}\equiv \left( q_{0}+ip^{0},q_{i}-ip^{i}\right) ^{T},
\label{X-def}
\end{equation}
such that (\ref{Z-CP^n}) and (\ref{DiZ-CP^n}) can be recast in the following
simple form:
\begin{eqnarray}
Z &=&e^{\mathcal{K}/2}\left( \mathcal{X}^{0}+\mathcal{X}^{i}t^{i}\right) ;
\label{Z-CP^n-X} \\
Z_{i} &=&2e^{3\mathcal{K}/2}\left( \frac{1}{2}e^{-\mathcal{K}}\mathcal{X}%
^{i}+\mathcal{X}^{0}\overline{t}^{\overline{i}}+\mathcal{X}^{j}t^{j}%
\overline{t}^{\overline{i}}\right) .  \label{DiZ-CP^n-X}
\end{eqnarray}

In the basis in which the charges $\mathcal{Q}$ or $\mathcal{X}$ are dressed
by the scalar fields into the central charge $Z$ and its K\"{a}hler
covariant derivatives $Z_{i}$, the quadratic invariant $\mathcal{I}_{2}$ of
the symplectic representation $\mathbf{1+n}$ of the electric-magnetic duality%
\footnote{%
We will henceforth simply refer to electric-magnetic duality as to duality.
In string theory, electric-magnetic duality can be seen as the
``continuous'' version, valid for large values of the charges, of the $U$%
-duality \cite{HT}.} group $U\left( 1,n\right) $ reads \cite
{ADF-U-duality-d=4}
\begin{equation}
\mathcal{I}_{2}=2\left( Z\overline{Z}-g^{i\overline{j}}Z_{i}\overline{Z}_{%
\overline{j}}\right) ,
\end{equation}
where $g_{i\overline{j}}\equiv \overline{\partial }_{\overline{j}}\partial
_{i}\mathcal{K}$ is the metric of the scalar manifold.

The BH effective potential and its criticality equations (\textit{alias}
Attractor Eqs. \cite{AM-Refs}) respectively read \cite{FGK}
\begin{eqnarray}
V &=&Z\overline{Z}+g^{i\overline{j}}Z_{i}\overline{Z}_{\overline{j}}; \\
\partial _{i}V &=&2\overline{Z}Z_{i}=0.  \label{CP^n-AEs}
\end{eqnarray}
The solutions to (\ref{CP^n-AEs}), such that $\left. V\right| _{\partial
_{i}V=0}\neq 0$ and its Hessian is positive definite, correspond to the
various classes of attractors\footnote{%
In the non-BPS case $\partial _{i}V=0$ corresponds to only one complex
equation ($Z=0$). Thus, a complex $\left( n-1\right) $-dimensional \textit{%
``moduli space''} of attractor solutions (namely the manifold $\mathbb{CP}%
^{n-1}$) exists in this case \cite{Ferrara-Marrani-2}.} in the BH
near-horizon geometry\footnote{%
The subscript ``$H$'' denotes evaluation at the BH horizon throughout.}:
\begin{eqnarray}
\text{BPS} &:&Z_{H}\neq 0,~Z_{i,H}=0~\forall i;~\mathcal{I}_{2}\left(
\mathcal{Q}\right) >0;  \label{pa1} \\
\text{non-BPS} &:&Z_{H}=0,~Z_{i,H}\neq 0~\text{for~some~}i;~\mathcal{I}%
_{2}\left( \mathcal{Q}\right) <0.  \label{pa2}
\end{eqnarray}
The attractor configurations are usually named ``large'', because they
correspond, through the Bekenstein-Hawking entropy-area formula \cite{BH},
to a non-vanishing (semi-)classical BH entropy given by (\ref{first}).

In the \textit{minimally coupled} models under consideration, there is also
a class of charge configurations supporting ``small'' single-center BHs
(which are BPS) with $\mathcal{I}_{2}=0$. Note that in Eqs. (\ref{pa1}) and (%
\ref{pa2}) $\mathcal{Q}$ is assumed to support single-center solutions.
Within the same assumption, note that
\begin{eqnarray}
\mathcal{I}_{2}\left( \mathcal{Q}\right) &>&0\Rightarrow Z\left( \mathcal{Q}%
\right) \neq 0; \\
\mathcal{I}_{2}\left( \mathcal{Q}\right) &<&0\Rightarrow D_{i}Z\left(
\mathcal{Q}\right) \neq 0~\text{for~some~}i.
\end{eqnarray}
Thus, as mentioned above, the \textit{minimally coupled} models have the
remarkable feature that the BPS (non-BPS) scalar flow trees never cross
points at which $Z=0$ ($D_{i}Z=0~\forall i$), due to the very constraints on
the supporting charge vectors.\medskip

Considering two different symplectic charge vectors
\begin{equation}
\mathcal{Q}_{1}\equiv \left( p^{0},p^{i},q_{0},q_{i}\right) ^{T};~\mathcal{Q}%
_{2}\equiv \left( P^{0},P^{i},Q_{0},Q_{i}\right) ^{T},~  \label{2-vectors}
\end{equation}
all the quadratic $U\left( 1,n\right) $-invariants built out with $\mathcal{Q%
}_{1}$ and $\mathcal{Q}_{2}$ read as follows\footnote{%
Note that we adopt a different normalization of $\mathcal{I}_{2}$ with
respect to \cite{Gnecchi-1}.
\par
Moreover, the subscripts ``$s$'' and ``$a$'' respectively stand for
``symmetric'' and ``antisymmetric'' with respect to the exchange $\mathcal{Q}%
_{1}\leftrightarrow \mathcal{Q}_{2}$.}:
\begin{eqnarray}
\mathbf{I}_{1} &\equiv &\mathcal{I}_{2}\left( \mathcal{Q}_{1}\right) =\left(
p^{0}\right) ^{2}-\left( p^{i}\right) ^{2}+q_{0}^{2}-q_{i}^{2};  \label{1} \\
\mathbf{I}_{2} &\equiv &\mathcal{I}_{2}\left( \mathcal{Q}_{2}\right) =\left(
P^{0}\right) ^{2}-\left( P^{i}\right) ^{2}+Q_{0}^{2}-Q_{i}^{2};  \label{2} \\
\mathbf{I}_{s} &\equiv &p^{0}P^{0}-p^{i}P^{i}+q_{0}Q_{0}-q_{i}Q_{i};
\label{3} \\
\mathbf{I}_{a} &\equiv
&p^{0}Q_{0}+p^{i}Q_{i}-q_{0}P^{0}-q_{i}P^{i}=-\left\langle \mathcal{Q}_{1},%
\mathcal{Q}_{2}\right\rangle .  \label{4}
\end{eqnarray}
In the complex parametrization of the charge vectors, (\ref{2-vectors})
amounts to considering
\begin{equation}
\mathcal{X}_{1}\equiv \left( q_{0}+ip^{0},q_{i}-ip^{i}\right) ^{T};~\mathcal{%
X}_{2}\equiv \left( Q_{0}+iP^{0},Q_{i}-iP^{i}\right) ^{T},  \label{X}
\end{equation}
and thus the four quadratic $U\left( 1,n\right) $-invariants (\ref{1})-(\ref
{4}) can be re-written as follows:
\begin{eqnarray}
\mathbf{I}_{1} &=&\mathcal{X}_{1}\cdot \overline{\mathcal{X}_{1}};
\label{r-1} \\
\mathbf{I}_{2} &=&\mathcal{X}_{2}\cdot \overline{\mathcal{X}_{2}}; \\
\mathbf{I}_{s} &=&\text{Re}\left( \mathcal{X}_{1}\cdot \overline{\mathcal{X}%
_{2}}\right) ; \\
\mathbf{I}_{a} &=&\text{Im}\left( \mathcal{X}_{1}\cdot \overline{\mathcal{X}%
_{2}}\right) ,  \label{r-4}
\end{eqnarray}
where ``$\cdot $'' is the bilinear Hermitian form defined by the Lorentzian
metric
\begin{equation}
\eta _{\Lambda \Sigma }=\text{diag}\left( 1,\overset{n}{\overbrace{-1,...,-1}%
}\right) ,
\end{equation}
namely:
\begin{eqnarray}
\mathcal{X}_{1}\cdot \overline{\mathcal{X}_{1}} &\equiv &\mathcal{X}%
_{1}^{\Lambda }\overline{\mathcal{X}_{1}^{\Sigma }}\eta _{\Lambda \Sigma };
\\
\mathcal{X}_{1}\cdot \overline{\mathcal{X}_{2}} &\equiv &\mathcal{X}%
_{1}^{\Lambda }\overline{\mathcal{X}_{2}^{\Sigma }}\eta _{\Lambda \Sigma }.
\label{deff}
\end{eqnarray}

From the expression (\ref{Z-CP^n-X}), it is easy to see that $Z$ transforms
as
\begin{equation}
Z\longrightarrow Ze^{i\alpha }
\end{equation}
under
\begin{equation}
\left\{
\begin{array}{l}
\mathcal{X}\longrightarrow \mathcal{X}e^{i\alpha }; \\
t^{i}\longrightarrow t^{i},
\end{array}
\right.
\end{equation}
namely a finite transformation of the \textit{global} (inactive on scalar
fields) $U\left( 1\right) $ factor of the duality group $U\left( 1,n\right)
=U(1)\times SU\left( 1,n\right) $. Such a $U\left( 1\right) $ is a global
electric-magnetic duality, which enlarges the actual duality group from the
numerator group $SU\left( 1,n\right) $ of the (non-compact) $\mathbb{CP}^{n}$
scalar manifold to $U\left( 1,n\right) $. Note that this is consistent also
with the fact that in the $n=0$ case of \textit{minimal coupling} sequence
(corresponding to ``pure'' $\mathcal{N}=2$ supergravity), the resulting
duality group is $U\left( 1\right) $.

In the one-modulus case, the presence of the global $U\left( 1\right) $
factor in the duality group can also be understood by noticing that a
consistent truncation of \textit{``pure''} $\mathcal{N}=4$ supergravity
produces the $n=1$ \textit{minimally coupled} $\mathcal{N}=2$ model in the
so-called \textit{axion-dilaton} symplectic basis (which is not the one
considered in\ Sec. \ref{n=1}; see \textit{e.g.} \cite{FHM-rev} for a recent
review and a list of Refs.). At the level of duality group, the
aforementioned truncation amounts to the following group embedding:
\begin{equation}
\underset{\mathcal{N}=4\text{,~}d=4\text{~``pure''}}{SL\left( 2,\mathbb{R}%
\right) \times SO\left( 6\right) }\supset \underset{\mathcal{N}=2\text{,~}d=4%
\text{~\textit{axion-dilaton}}}{SL\left( 2,\mathbb{R}\right) \times SO\left(
2\right) }.
\end{equation}
Notice that the axion-dilaton of the resulting \textit{minimally coupled} $%
\mathcal{N}=2$ theory is nothing but the axio-dilatonic scalar of the $%
\mathcal{N}=4$ supergravity multiplet. Moreover, four of the six $\mathcal{N}%
=2$ graviphotons are truncated away, and the remaining two ones split into
the $\mathcal{N}=2$ graviphoton and in the Maxwell field of the
axio-dilatonic $\mathcal{N}=2$ multiplet. At fermionic level, two out of the
four $\mathcal{N}=4$ gravitinos are truncated away, consistent with the
lower local supersymmetry. The supersymmetry uplift of $\mathcal{N}=2$
axion-dilaton model into extended supergravities has been recently discussed
\textit{e.g.} in \cite{Gnecchi-2}.

It is worth remarking that without the extra global $U\left( 1\right) $ in
the duality group, the analysis that we are going to perform in Sec. \ref
{n=1} would have been incomplete. Indeed, from their very definitions, $%
\mathbf{I}_{1}$, $\mathbf{I}_{2}$, $\mathbf{I}_{s}$ and $\mathbf{I}_{a}$ are
two-center invariants of both $SU\left( 1,1\right) $ and $U\left( 1,1\right)
$. However, $SU\left( 1,1\right) $ has an extra two-center quadratic
invariant, defined as\footnote{%
Note that, from its very definition (\ref{I-Fraktur}), $\frak{I}$ exists for
the duality group $SU\left( 1,n\right) $ only when $n+1$ centers are
considered (this statement holds irrespective of the non-compact nature of
the duality group itself).}
\begin{equation}
\frak{I}\equiv \mathcal{X}_{1}\wedge \mathcal{X}_{2}\equiv \mathcal{X}%
_{1}^{\Lambda }\mathcal{X}_{2}^{\Sigma }\epsilon _{\Lambda \Sigma },
\label{I-Fraktur}
\end{equation}
where $\epsilon $ denotes the antisymmetric Levi-Civita symbol. Note that $%
\frak{I}$ is the unique quadratic two-center $SU\left( 1,1\right) $%
-invariant which is complex, and thus which is not an $U\left( 1,1\right) $%
-invariant. Its squared absolute value is related to $\mathbf{I}_{1}$, $%
\mathbf{I}_{2}$, $\mathbf{I}_{s}$ and $\mathbf{I}_{a}$ as follows:
\begin{equation}
\left| \frak{I}\right| ^{2}=-\mathbf{I}_{1}\mathbf{I}_{2}+\mathbf{I}_{s}^{2}+%
\mathbf{I}_{a}^{2}.  \label{rell}
\end{equation}
%If this were not the case, one would have also had to consider it as an
%independent \textit{quartic} two-center $U\left( 1,1\right) $-invariant.
%But, s
Since (\ref{rell}) holds, $\frak{I}$ would only have introduced a further
real degrees of freedom (charge) %, corresponding to its argument
%\begin{equation}
%\arg \left( \frak{I}\right) =-i\ln \left( \frac{\frak{I}}{\sqrt{-\mathbf{I}%
%_{1}\mathbf{I}_{2}+\mathbf{I}_{s}^{2}+\mathbf{I}_{a}^{2}}}\right) ,
%\end{equation}
in the discussion of Sec. \ref{n=1}.\medskip

The following relation will prove to be useful in the treatment given below:
\begin{gather}
\mathcal{I}_{2}\left( \mathcal{Q}_{1}+\mathcal{Q}_{2}\right) =\mathbf{I}_{1}+%
\mathbf{I}_{2}+2\mathbf{I}_{s};  \label{I2(Q1+Q2)} \\
\Downarrow  \notag \\
\mathcal{I}_{2}\left( \mathcal{Q}_{1}+\mathcal{Q}_{2}\right) \gtreqless
0\Leftrightarrow \mathbf{I}_{s}\gtreqless -\frac{1}{2}\left( \mathbf{I}_{1}+%
\mathbf{I}_{2}\right) ; \\
\mathcal{I}_{2}\left( \mathcal{Q}_{1}+\mathcal{Q}_{2}\right) \gtreqless
\mathbf{I}_{1}+\mathbf{I}_{2}\Leftrightarrow \mathbf{I}_{s}\gtreqless 0.
\end{gather}
In particular, it holds that
\begin{equation}
\mathbf{I}_{s}>0\Rightarrow \mathcal{I}_{2}\left( \mathcal{Q}_{1}+\mathcal{Q}%
_{2}\right) >\mathbf{I}_{1}+\mathbf{I}_{2}.  \label{rel-1}
\end{equation}

In the case of two-center BH solutions with both BPS centers (\textit{i.e.} $%
\mathbf{I}_{1}\geqslant 0$ and $\mathbf{I}_{2}\geqslant 0$), in Secs. \ref
{Axion-Dilaton-BPS} and \ref{n>1-BPS} we will obtain, in terms of the
aforementioned duality-invariants, the stability region $\mathcal{S}$ of the
composite solution, and the MS region (\textit{if any}). In fact, depending
on the sign of some invariants, we have found that MS or AMS walls can occur
in the scalar manifold, \textit{but not both}. Therefore, the scalar flow
supported by the physical charge orbit, whose stability region ends when
crossing the MS wall, never encounters the AMS wall, which instead pertains
to another (un-physical) charge orbit which does not support a MS wall.

More interestingly, under the assumption of existence of a MS wall (and of a
stability region of the two-center solution), we have found that $\mathcal{I}%
_{2}\left( \mathcal{Q}_{1}+\mathcal{Q}_{2}\right) >0$, and in particular
that (\ref{rel-1}) necessarily holds. This latter, through (\ref{first}),
leads to the following fundamental relation (\ref{EntropyBound}) anticipated
above:
\begin{equation}
S_{1\text{-ctr,BPS}}\left( \mathcal{Q}_{1}+\mathcal{Q}_{2}\right) >S_{2\text{%
-ctr,BPS}}\left( \mathcal{Q}_{1},\mathcal{Q}_{2}\right) =S_{1\text{-ctr,BPS}%
}\left( \mathcal{Q}_{1}\right) +S_{1\text{-ctr,BPS}}\left( \mathcal{Q}%
_{2}\right) ,  \label{PA-2}
\end{equation}
namely that the entropy of the single-center solution with charge $\mathcal{Q%
}_{1}+\mathcal{Q}_{2}$ is always larger than the entropy of the two-center
solution with charges $\mathcal{Q}_{1}$ and $\mathcal{Q}_{2}$ for the
centers $1$ and $2$, respectively. This can ultimately be traced back to the
fact that the BH entropy (\ref{first}) can be written as
\begin{equation}
S\left( \mathcal{Q}\right) =\frac{\pi }{2}\mathcal{X}\cdot \overline{%
\mathcal{X}},  \label{bek-2}
\end{equation}
and that the MS condition requires
\begin{equation}
\mathbf{I}_{s}>0,
\end{equation}
from which (\ref{rel-1}) and (\ref{PA-2}) follow.

\textit{Mutatis mutandis}, the same holds for two-center BH solutions with
both non-BPS centers (\textit{i.e.} $\mathbf{I}_{1}<0$ and $\mathbf{I}_{2}<0$%
), at least in the \textit{minimally coupled} model with $n=1$ complex
scalar. In such a framework, it holds
\begin{equation}
\mathbf{I}_{s}<0\Rightarrow \mathcal{I}_{2}\left( \mathcal{Q}_{1}+\mathcal{Q}%
_{2}\right) <\mathbf{I}_{1}+\mathbf{I}_{2}=-\left| \mathbf{I}_{1}+\mathbf{I}%
_{2}\right| .  \label{rel-2}
\end{equation}
In Sec. \ref{Axion-Dilaton-nBPS} we will show that the assumption of
existence of a MS wall (and of a stability region of the two-center
solution) necessarily implies $\mathcal{I}_{2}\left( \mathcal{Q}_{1}+%
\mathcal{Q}_{2}\right) <0$, and in particular (\ref{rel-2}). This latter,
through the formula (\ref{first}), implies
\begin{equation}
S_{1\text{-ctr,nBPS}}\left( \mathcal{Q}_{1}+\mathcal{Q}_{2}\right) >S_{2%
\text{-ctr,nBPS}}\left( \mathcal{Q}_{1},\mathcal{Q}_{2}\right) =S_{1\text{%
-ctr,nBPS}}\left( \mathcal{Q}_{1}\right) +S_{1\text{-ctr,nBPS}}\left(
\mathcal{Q}_{2}\right) .  \label{PA-2-2}
\end{equation}
The treatment of the non-BPS case is possible in virtue of the observation
\cite{MS-FM-1} that the ``fake'' superpotential \cite{Ceresole:2007wx,ADOT-1}%
, which gives the non-BPS ADM mass, also satisfies a Cauchy-Schwarz triangle
inequality:
\begin{equation}
W\left( \mathcal{Q}_{1}+\mathcal{Q}_{2}\right) \leqslant W\left( \mathcal{Q}%
_{1}\right) +W\left( \mathcal{Q}_{2}\right) ,  \label{TriangleIneq}
\end{equation}
as it holds for the central charge $Z$ in the BPS case.

Eqs. (\ref{EntropyBound}), (\ref{PA-2}) and (\ref{PA-2-2}) express an
interesting feature of the \textit{minimally coupled} models of $\mathcal{N}%
=2$, $d=4$ Maxwell-Einstein supergravity: the constituents always have an
entropy which is smaller than the entropy of the original composite (if
considered as a single-center solution). Thus, the corresponding split
dynamics of the scalar flows exhibits a different behavior with respect to
the $\mathcal{N}=2$ models with special K\"{a}hler geometry based on cubic
prepotentials. Indeed, in these latter models, MS and AMS walls are known to
co-exist, for a suitable choice of the charge vectors $\mathcal{Q}_{1}$ and $%
\mathcal{Q}_{2}$, in different zones of the scalar manifold itself (see
\textit{e.g.} \cite{ADMJ-2}, and the analysis in Sec. \ref{BPS-t^3-Analysis}
\cite{BD-1}). Furthermore, in cubic $\mathcal{N}=2$ models, also by assuming
$\mathcal{I}_{4}\left( \mathcal{Q}_{1}\right) \geqslant 0$ and $\mathcal{I}%
_{4}\left( \mathcal{Q}_{2}\right) \geqslant 0$, $\mathcal{I}_{4}\left(
\mathcal{Q}_{1}+\mathcal{Q}_{2}\right) $ is not necessarily positive (see
\textit{e.g.} \cite{BD-1} and \cite{Castro-Simon}).

Eqs.(\ref{PA-2}) and (\ref{PA-2-2}) also imply that ``entropy
enigma'' decays \cite{DM-1,DM-2,David} never occur in these models,
and thus that in the corresponding regime of large charges the
microscopic state counting is still dominated by the single-center
configurations (see \textit{e.g.} the discussion in
\cite{DM-1,DM-2}).

As mentioned in Sec. \ref{Intro}, Eqs. (\ref{bek-2}) and (\ref{TriangleIneq}%
) also imply that the BPS (non-BPS) mass is bounded from below by the
single-center entropy
\begin{equation}
\left| Z\left( t^{i}\left( r\right) ,\overline{t}^{\overline{i}}\left(
r\right) ;\mathcal{Q}_{1}+\mathcal{Q}_{2}\right) \right| \geqslant \sqrt{%
\frac{S_{1\text{-ctr,BPS}}\left( \mathcal{Q}_{1}+\mathcal{Q}_{2}\right) }{%
\pi }}=\sqrt{\frac{\mathcal{I}_{2}\left( \mathcal{Q}_{1}+\mathcal{Q}%
_{2}\right) }{2}},  \label{ti-Z}
\end{equation}
\begin{equation}
W\left( t^{i}\left( r\right) ,\overline{t}^{\overline{i}}\left( r\right) ;%
\mathcal{Q}_{1}+\mathcal{Q}_{2}\right) \geqslant \sqrt{\frac{S_{1\text{%
-ctr,nBPS}}\left( \mathcal{Q}_{1}+\mathcal{Q}_{2}\right) }{\pi }}=\sqrt{-%
\frac{\mathcal{I}_{2}\left( \mathcal{Q}_{1}+\mathcal{Q}_{2}\right) }{2}}\,.
\label{ti-W}
\end{equation}

As a consequence of (\ref{rel-1})-(\ref{PA-2-2}), the inequality (\ref{ti-Z}%
) (and its non-BPS counterpart (\ref{ti-W})) implies that $Z$ ($W$) never
vanishes in the scalar manifold, neither for single-center nor for
two-center solutions. For this reason, and for the fact that MS and AMS
walls cannot co-exist in the scalar manifold, the ``paradox'' which led to
the introduction of ``bound state transformation walls'' \cite{ADMJ-2} does
not occur in the class of theories under consideration.

It is worth recalling that the so-called $t^{2}$ and $t^{3}$ models are the
only one-modulus $\mathcal{N}=2$, $d=4$ Maxwell-Einstein supergravity models
with homogeneous scalar manifolds \cite{dWVVP}. As mentioned above, in cubic
special geometries ``bound state recombination walls'' and ``entropy
enigma'' decays are possible, respectively because (\ref{ti-Z}) (with $%
\mathcal{I}_{2}$ replaced by $\mathcal{I}_{4}$) and (\ref{PA-2}) do not
necessarily apply.

\section{\label{n=1}One Modulus}

We start and consider the simplest model, namely the one with $n=1$ \textit{%
minimally coupled} vector multiplet, duality-related to the so-called
\textit{axion-dilaton} model (see \textit{e.g.} \cite{FHM-rev} for a review
and a list of Refs.). The metric function in this case reads:
\begin{eqnarray}
g^{t\overline{t}} &=&\left( 1-t\overline{t}\right) ^{2}=\left| e_{\widehat{t}%
}^{t}\right| ^{2}; \\
e_{\widehat{t}}^{t} &=&i\left( 1-t\overline{t}\right) ,
\end{eqnarray}
where the phase of the \textit{Vielbein} $e_{\widehat{t}}^{t}$ is chosen for
later convenience.

The domain of definition of the K\"{a}hler potential $\mathcal{K}$ and of
the metric $g^{t\overline{t}}$ is the open unit disc centered in the origin
of the Argand-Gauss plane (we use the notation $b\equiv $Re$\left( t\right) $
and $a\equiv $Im$\left( t\right) $):
\begin{equation}
b^{2}+a^{2}<1.  \label{metric-domain}
\end{equation}

The expressions of the central charge and of the \textit{matter charge} are
given by the $n=1$ case of Eqs. (\ref{Z-CP^n})-(\ref{DiZ-CP^n}), whereas the
BPS and non-BPS attractor values of the complex scalar $t$ respectively read
as follows \cite{Gnecchi-1}:
\begin{eqnarray}
t_{BPS} &=&-\frac{\left( q_{1}+ip^{1}\right) }{\left( q_{0}-ip^{0}\right) };
\label{t-BPS} \\
t_{nBPS} &=&-\frac{\left( q_{0}+ip^{0}\right) }{\left( q_{1}-ip^{1}\right) }.
\label{t-nBPS}
\end{eqnarray}

\subsection{\label{Axion-Dilaton-BPS}BPS MS \textit{or} AMS Wall}

Within this Subsection, we assume
\begin{equation}
\left( \mathcal{Q}_{1},\mathcal{Q}_{2}\right) :\left\{
\begin{array}{l}
\mathbf{I}_{1}\geqslant 0; \\
\\
\mathbf{I}_{2}\geqslant 0,
\end{array}
\right.  \label{BPS-charges}
\end{equation}
as well as $\mathbb{CP}^{1}$ to be the \textit{spatially asymptotical} ($%
r\rightarrow \infty $) scalar manifold.

Depending on the various cases, the \textit{a priori} possible BPS ``large''
two-center configurations are\footnote{%
Throughout the present paper, we consider only ``large'' initial states.
From the reasonings done in Sec. \ref{Intro} and the main results of the
present investigation, when requiring the existence of a stability region
and of a MS wall, this assumption does not imply any loss of generality.}
\begin{eqnarray}
\mathbf{1}.~\text{BPS~``large''} &\rightarrow &\text{BPS~``large''}+\text{%
BPS~``large''};  \label{1-BPS} \\
\mathbf{2}.~\text{BPS~``large''} &\rightarrow &\text{BPS~``large''}+\text{%
BPS~``small''};  \label{2-BPS} \\
\mathbf{3}.~\text{BPS~``large''} &\rightarrow &\text{BPS~``small''}+\text{%
BPS~``small''}.  \label{3-BPS}
\end{eqnarray}

The BPS MS and AMS walls are defined as (within $\mathbb{CP}^{1}$;we use the
notation $Z_{\mathbf{a}}\equiv Z\left( b,a;\mathcal{Q}_{\mathbf{a}}\right) $%
, $\mathbf{a}=1,2$ throughout):
\begin{eqnarray}
MS_{BPS} &\equiv &\left\{ b+ia:\left[
\begin{array}{l}
\text{Im}\left( Z_{1}\overline{Z_{2}}\right) =0; \\
\\
\text{Re}\left( Z_{1}\overline{Z_{2}}\right) >0;
\end{array}
\right. \right\} ;  \label{MS-BPS} \\
&&  \notag \\
AMS_{BPS} &\equiv &\left\{ b+ia:\left[
\begin{array}{l}
\text{Im}\left( Z_{1}\overline{Z_{2}}\right) =0; \\
\\
\text{Re}\left( Z_{1}\overline{Z_{2}}\right) <0;
\end{array}
\right. \right\} ,  \label{AMS-BPS}
\end{eqnarray}
where
\begin{eqnarray}
2\left( 1-b^{2}-a^{2}\right) \text{Re}\left( Z_{1}\overline{Z_{2}}\right) &=&%
\left[
\begin{array}{l}
q_{0}Q_{0}+p^{0}P^{0} \\
+\left( q_{1}Q_{1}+p^{1}P^{1}\right) b^{2}+\left(
q_{0}Q_{1}+q_{1}Q_{0}-p^{0}P^{1}-p^{1}P^{0}\right) b \\
+\left( q_{1}Q_{1}+p^{1}P^{1}\right) a^{2}+\left(
q_{0}P^{1}+q_{1}P^{0}+p^{0}Q_{1}+p^{1}Q_{0}\right) a
\end{array}
\right] ;  \notag \\
&&  \label{Y-call} \\
2\left( 1-b^{2}-a^{2}\right) \text{Im}\left( Z_{1}\overline{Z_{2}}\right) &=&%
\left[
\begin{array}{l}
p^{0}Q_{0}-q_{0}P^{0} \\
+\left( q_{1}P^{1}-p^{1}Q_{1}\right) b^{2}+\left(
q_{0}P^{1}-q_{1}P^{0}+p^{0}Q_{1}-p^{1}Q_{0}\right) b \\
+\left( q_{1}P^{1}-p^{1}Q_{1}\right) a^{2}+\left(
-q_{0}Q_{1}+q_{1}Q_{0}+p^{0}P^{1}-p^{1}P^{0}\right) a
\end{array}
\right] ;  \notag \\
&&  \label{X-call}
\end{eqnarray}

The region of stability $\mathcal{S}_{BPS}$ of the two-center BPS solution
is defined as
\begin{equation}
\mathcal{S}_{BPS}\equiv \left\{ b+ia\in \mathbb{CP}^{1}:\left\langle
\mathcal{Q}_{1},\mathcal{Q}_{2}\right\rangle \text{Im}\left( Z_{1}\overline{%
Z_{2}}\right) >0\right\} .  \label{S-BPS}
\end{equation}

The distance between the centers $1$ and $2$ can be $SU\left( 1,1\right) $%
-invariantly written as \cite{BD-1}
\begin{equation}
\left| \overrightarrow{x}_{1}-\overrightarrow{x}_{2}\right| =\frac{%
\left\langle \mathcal{Q}_{1},\mathcal{Q}_{2}\right\rangle }{2}\frac{\left|
Z_{1}+Z_{1}\right| }{\text{Im}\left( Z_{1}\overline{Z_{2}}\right) },
\label{BPS-distance}
\end{equation}
and the corresponding configurational angular momentum reads \cite{D-1,BD-1}
\begin{equation}
\overrightarrow{J}=\frac{\left\langle \mathcal{Q}_{1},\mathcal{Q}%
_{2}\right\rangle }{2}\frac{\left( \overrightarrow{x}_{1}-\overrightarrow{x}%
_{2}\right) }{\left| \overrightarrow{x}_{1}-\overrightarrow{x}_{2}\right| }=%
\frac{\text{Im}\left( Z_{1}\overline{Z_{2}}\right) }{\left|
Z_{1}+Z_{1}\right| }\left( \overrightarrow{x}_{1}-\overrightarrow{x}%
_{2}\right) .  \label{BPS-J}
\end{equation}

It is also here worth observing that the ``large'' BPS single-center
solution with charge $\mathcal{Q}=\mathcal{Q}_{1}+\mathcal{Q}_{2}$ would
exist \textit{iff}
\begin{equation}
\mathcal{I}_{2}\left( \mathcal{Q}_{1}+\mathcal{Q}_{2}\right)
>0\Leftrightarrow 2\mathbf{I}_{s}>-\left( \mathbf{I}_{1}+\mathbf{I}%
_{2}\right) ,  \label{large-BPS-single-cond}
\end{equation}
where (\ref{I2(Q1+Q2)}) has been used, and the condition (\ref{BPS-charges})
must be taken into account.

\subsubsection{\label{BPS-Case-1}Case \textbf{1}}

The most general charge configuration supporting the two-center solution (%
\ref{1-BPS}) is duality-related to\footnote{%
We remind that at the attractor points $2$Im$\left( Z_{1}\bar{Z}_{2}\right)
=-\langle \mathcal{Q}_{1},\mathcal{Q}_{2}\rangle $ as pointed out in \cite
{D-1,MS-FM-1}. It turns out that this relation still holds in our case at
the single center attractor point with charge $\mathcal{Q}_{1}+\mathcal{Q}%
_{2}$.
\par
Furthermore, by using the fundamental identities of special K\"{a}hler
geometry in presence of two symplectic charge vectors $\mathcal{Q}_{1}$ and $%
\mathcal{Q}_{2}$ (see \textit{e.g.} \cite{D-1,BFM-1,FK-N=8}), one can
compute that at BPS attractor points for the centers $1$ or $2$:
\begin{equation*}
\text{Re}\left( Z_{1}\bar{Z}_{2}\right) =-\frac{1}{2}\mathcal{Q}_{1}^{T}%
\mathcal{MQ}_{2},
\end{equation*}
where $\mathcal{M}$ is the symplectic, symmetric, negative definite $2\left(
n_{V}+1\right) \times 2\left( n_{V}+1\right) $ matrix with entries depending
on the real and imaginary part of the vector kinetic matrix $\mathcal{N}%
_{\Lambda \Sigma }$ (see \textit{e.g. }\cite{ADF-central,CDF-rev}, and Refs.
therein). Notice that $\mathcal{Q}_{1}^{T}\mathcal{MQ}_{2}$ does not have a
definite sign.}
\begin{eqnarray}
\mathcal{Q}_{1} &\equiv &\left( 0,0,q_{0},0\right) \Rightarrow \left\{
\begin{array}{l}
\mathbf{I}_{1}=q_{0}^{2}>0; \\
\\
t_{H,BPS}\left( \mathcal{Q}_{1}\right) =0;
\end{array}
\right.   \label{Q1-BPS-large} \\
&&  \notag \\
\mathcal{Q}_{2} &\equiv &\left( P^{0},P^{1},Q_{0},0\right) \Rightarrow
\left\{
\begin{array}{l}
\mathbf{I}_{2}=\left( P^{0}\right) ^{2}+Q_{0}^{2}-\left( P^{1}\right) ^{2}>0;
\\
\\
t_{H,BPS}\left( \mathcal{Q}_{2}\right) =-i\frac{P^{1}}{\left(
Q_{0}-iP^{0}\right) },
\end{array}
\right.   \label{Q2-BPS-large-cond}
\end{eqnarray}
which can thus be considered without any loss in generality. Indeed, for the
charge configuration (\ref{Q1-BPS-large})-(\ref{Q2-BPS-large-cond}), the
four quadratic $U\left( 1,1\right) $-invariants (\ref{1})-(\ref{4}) are all
generally non-coinciding and non-vanishing:
\begin{equation}
\left( \mathcal{Q}_{1},\mathcal{Q}_{2}\right) :\left\{
\begin{array}{l}
\mathbf{I}_{1}=q_{0}^{2}; \\
\mathbf{I}_{2}=\left( P^{0}\right) ^{2}-\left( P^{1}\right) ^{2}+Q_{0}^{2}>0;
\\
\mathbf{I}_{s}=q_{0}Q_{0}; \\
\mathbf{I}_{a}=-q_{0}P^{0}.
\end{array}
\right.   \label{n=1-BPS-invs}
\end{equation}
It is worth noting that the charge vector $\mathcal{Q}_{1}$ given by Eq. (%
\ref{Q1-BPS-large}), in which only $q_{0}$ is non-vanishing, is nothing but
the Reissner-Nordstr\"{o}m black hole embedded in $\mathbb{C\mathbb{P}}^{1}$%
, with attractor value at the origin of such a space.

A manifestly $U\left( 1,1\right) $-invariant characterization of the four
non-vanishing charges of the general BPS two-center configuration (\ref
{Q1-BPS-large})-(\ref{Q2-BPS-large-cond}) reads as follows:
\begin{eqnarray}
q_{0}^{2} &=&\mathbf{I}_{1};  \label{q0-inv} \\
\left( P^{0}\right) ^{2} &=&\frac{\mathbf{I}_{a}^{2}}{\mathbf{I}_{1}};
\label{P0-inv} \\
\left( P^{1}\right) ^{2} &=&\frac{\left( \mathbf{I}_{s}^{2}+\mathbf{I}%
_{a}^{2}-\mathbf{I}_{1}\mathbf{I}_{2}\right) }{\mathbf{I}_{1}};
\label{P1-inv} \\
Q_{0}^{2} &=&\frac{\mathbf{I}_{s}^{2}}{\mathbf{I}_{1}},  \label{Q0-inv}
\end{eqnarray}
where
\begin{equation}
\left.
\begin{array}{r}
\mathbf{I}_{1}>0 \\
\mathbf{I}_{2}>0
\end{array}
\right\} \Rightarrow \mathbf{I}_{s}^{2}+\mathbf{I}_{a}^{2}-\mathbf{I}_{1}%
\mathbf{I}_{2}>0.  \label{PA-1}
\end{equation}

Within the configuration (\ref{Q1-BPS-large})-(\ref{Q2-BPS-large-cond}), the
real and imaginary part of $Z_{1}\overline{Z_{2}}$ respectively read (recall
(\ref{Y-call}) and (\ref{X-call})):
\begin{eqnarray}
\text{Re}\left( Z_{1}\overline{Z}_{2}\right) &=&\frac{q_{0}\left(
Q_{0}+P^{1}a\right) }{2\left( 1-b^{2}-a^{2}\right) };  \label{Y-call-simpl}
\\
\text{Im}\left( Z_{1}\overline{Z}_{2}\right) &=&\frac{q_{0}\left(
-P^{0}+P^{1}b\right) }{2\left( 1-b^{2}-a^{2}\right) }.  \label{X-call-simpl}
\end{eqnarray}

\begin{figure}[h!]
\centering \epsfxsize= 10cm \epsfysize= 10cm \epsfbox{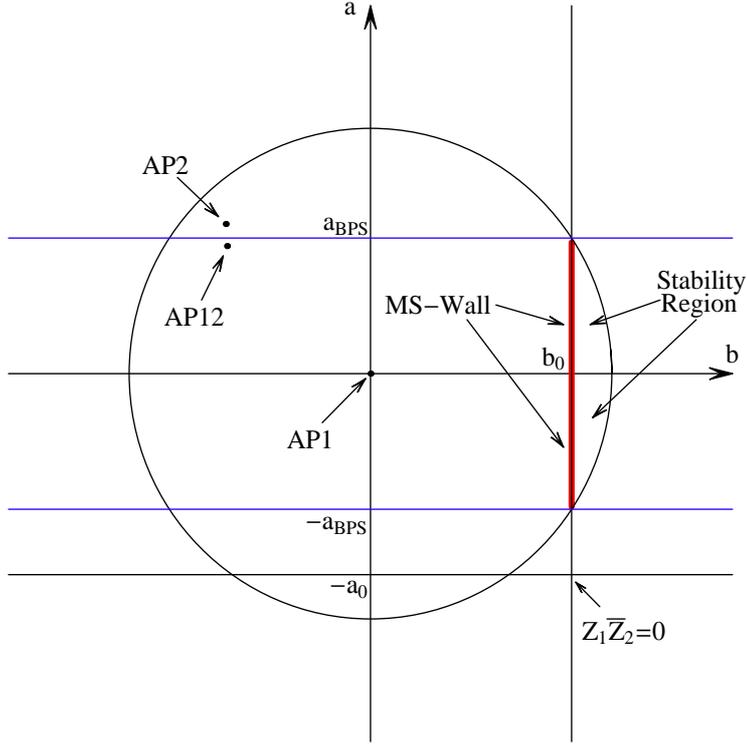}
\caption{Stability region $\mathcal{S}_{BPS}$ and MS wall $MS_{BPS}$ for the
BPS two-center extremal BH solution of $1$-modulus \textit{minimally coupled}
$\mathcal{N}=2$ model, represented as functions of $b$ and $a$, respectively
the real and imaginary part of the scalar $t$. The charges has been chosen
all positive. Here $b_0=P^0/P^1$ and $a_0=Q_0/P^1$ are respectively the
values at which Im$\left(Z_1\bar Z_2\right)$ and Re$\left(Z_1\bar Z_2\right)$
vanish. The attractor points associated to the centers with charges $%
\mathcal{Q}_1$, $\mathcal{Q}_2$ and $\mathcal{Q}_1+\mathcal{Q}_1$ are
respectively labeled by AP1, AP2 and AP12.}
\end{figure}

Let us start by computing the region of stability $\mathcal{S}_{BPS}$
defined in (\ref{S-BPS}):
\begin{equation}
\mathcal{S}_{BPS}:\frac{P^{1}}{P^{0}}b>1\Leftrightarrow \left\{
\begin{array}{l}
P^{0}P^{1}>0:\frac{P^{0}}{P^{1}}<b<\sqrt{1-a^{2}}; \\
\\
P^{0}P^{1}<0:-\sqrt{1-a^{2}}<b<\frac{P^{0}}{P^{1}}.
\end{array}
\right.  \label{S-BPS-gen}
\end{equation}
Note that $a$ enters Eq. (\ref{S-BPS-gen}) only through the constraint to
belong to the domain of definition of the metric of the scalar manifold,
defined by (\ref{metric-domain}):
\begin{equation}
b^{2}+a^{2}<1,  \label{metric-domain-x-y}
\end{equation}
implying that
\begin{equation}
\left| \frac{P^{0}}{P^{1}}\right| <1\Leftrightarrow \left( P^{1}\right)
^{2}-\left( P^{0}\right) ^{2}>0\Leftrightarrow \mathbf{I}_{s}^{2}-\mathbf{I}%
_{1}\mathbf{I}_{2}>0.  \label{impl}
\end{equation}
By using (\ref{P0-inv})-(\ref{P1-inv}), the region of stability $\mathcal{S}%
_{BPS}$ (\ref{S-BPS-gen})-(\ref{impl}) can be re-expressed as follows:
\begin{equation}
\mathcal{S}_{BPS}:\left\{
\begin{array}{l}
\pm \frac{\sqrt{\mathbf{I}_{s}^{2}+\mathbf{I}_{a}^{2}-\mathbf{I}_{1}\mathbf{I%
}_{2}}}{\left| \mathbf{I}_{a}\right| }b>1; \\
\\
b^{2}+a^{2}<1.
\end{array}
\right.  \label{S-BPS-gen-inv}
\end{equation}

Then, one can study the existence of the BPS MS and AMS walls, defined by (%
\ref{MS-BPS})-(\ref{Y-call}).

Within the condition (\ref{impl}), it is convenient to define (see Eqs. (\ref
{BPS-y-order})-(\ref{BPS-y-order-2}))
\begin{eqnarray}
a_{BPS} &\equiv &\sqrt{\frac{\left( P^{1}\right) ^{2}-\left( P^{0}\right)
^{2}}{\left( P^{1}\right) ^{2}}}=\sqrt{\frac{\mathbf{I}_{s}^{2}-\mathbf{I}%
_{1}\mathbf{I}_{2}}{\mathbf{I}_{s}^{2}+\mathbf{I}_{a}^{2}-\mathbf{I}_{1}%
\mathbf{I}_{2}}}>0;  \label{def-a_BPS} \\
&&  \notag \\
&&  \notag \\
\mathcal{A} &\equiv &\left\{ b,a\in \mathbb{CP}^{1}:\left[
\begin{array}{l}
b=\pm \frac{\left| \mathbf{I}_{a}\right| }{\sqrt{\mathbf{I}_{s}^{2}+\mathbf{I%
}_{a}^{2}-\mathbf{I}_{1}\mathbf{I}_{2}}}; \\
\\
-\left| \frac{Q_{0}}{P^{1}}\right| <-a_{BPS}<a<a_{BPS}<\left| \frac{Q_{0}}{%
P^{1}}\right| .
\end{array}
\right. \right\} ;  \label{def-A-call} \\
&&  \notag \\
\left| \frac{Q_{0}}{P^{1}}\right| &=&\frac{\left| \mathbf{I}_{s}\right| }{%
\sqrt{\mathbf{I}_{s}^{2}+\mathbf{I}_{a}^{2}-\mathbf{I}_{1}\mathbf{I}_{2}}}.
\end{eqnarray}
Then, through some straightforward computations (detailed in App. A), one
obtains that within the two-center charge configuration (\ref{Q1-BPS-large}%
)-(\ref{Q2-BPS-large-cond}) the existence of BPS MS or AMS walls depends on
the sign of $\mathbf{I}_{s}$:
\begin{eqnarray}
\mathbf{I}_{s} &>&0:\left\{
\begin{array}{l}
MS_{BPS}=\mathcal{A}; \\
\\
\nexists AMS_{BPS};
\end{array}
\right.  \label{q0Q0>0} \\
&&  \notag \\
\mathbf{I}_{s} &<&0:\left\{
\begin{array}{l}
\nexists MS_{BPS}; \\
\\
AMS_{BPS}=\mathcal{A}.
\end{array}
\right.  \label{q0Q0<0}
\end{eqnarray}
$\mathcal{S}_{BPS}$ and $MS_{BPS}$ are graphically depicted in Fig. 1 for an
all positive charge configuration.

\paragraph{Single-Center Solution and MS Wall}

By recalling (\ref{I2(Q1+Q2)}), it follows that
\begin{equation}
\mathbf{I}_{1}+\mathbf{I}_{2}+2\mathbf{I}_{s}>0  \label{I}
\end{equation}
is the general condition of existence of the ``large'' BPS single-center
solution with charge $\mathcal{Q}_{1}+\mathcal{Q}_{2}$. By denoting the
entropy of the BH solution with $S$, one then obtains that
\begin{equation}
S_{\text{1-ctr,BPS}}\left( \mathcal{Q}_{1}+\mathcal{Q}_{2}\right) \gtreqless
S_{\text{2-ctr,BPS}}\left( \mathcal{Q}_{1},\mathcal{Q}_{2}\right)
\Leftrightarrow \left\{
\begin{array}{l}
\mathbf{I}_{s}>0; \\
\mathbf{I}_{s}=0; \\
-\frac{1}{2}\left( \mathbf{I}_{1}+\mathbf{I}_{2}\right) <\mathbf{I}_{s}<0.
\end{array}
\right.  \label{S-ineq-BPS}
\end{equation}

As anticipated in Sec. \ref{Intro}, within the general conditions (\ref
{Q2-BPS-large-cond}) and (\ref{impl}) on $\mathcal{Q}_{2}$ (corresponding to
assuming the existence of a stability region for the two-center
configuration ``large'' BPS $+$ ``large'' BPS (\ref{1-BPS})), the existence
of a BPS MS wall $MS_{BPS}$ (see (\ref{q0Q0>0})) implies the existence of
the ``large'' BPS single-center solution with charge $\mathcal{Q}_{1}+%
\mathcal{Q}_{2}$, with entropy strictly larger than the entropy of the
two-center solution, as given by Eq. (\ref{PA-2}).

\subsubsection{\label{BPS-Case-2}Case \textbf{2}}

The most general charge configuration supporting the two-center solution (%
\ref{2-BPS}) is duality-related to
\begin{eqnarray}
\mathcal{Q}_{1} &\equiv &\left( 0,0,q_{0},0\right) \Rightarrow \left\{
\begin{array}{l}
\mathbf{I}_{1}=q_{0}^{2}>0; \\
t_{H,BPS}\left( \mathcal{Q}_{1}\right) =0;
\end{array}
\right.  \label{Q1-BPS-large-2} \\
&&  \notag \\
\mathcal{Q}_{2} &\equiv &\left( P^{0},P^{1},Q_{0},0\right) \Rightarrow
\left\{
\begin{array}{l}
\mathbf{I}_{2}=\left( P^{0}\right) ^{2}+Q_{0}^{2}-\left( P^{1}\right) ^{2}=0;
\\
\nexists t_{H}\left( \mathcal{Q}_{2}\right) ,
\end{array}
\right.  \label{Q2-BPS-small-2}
\end{eqnarray}
which can thus be considered without any loss in generality.

This case can be consistently obtained as the limit $\mathbf{I}%
_{2}\rightarrow 0^{+}$ of the treatment given in Sec. \ref{BPS-Case-1} and
in App. \ref{App-I}, enforcing the addition restriction
\begin{equation}
a_{BPS}=\left| \frac{Q_{0}}{P^{1}}\right| =\frac{\left| \mathbf{I}%
_{s}\right| }{\sqrt{\mathbf{I}_{s}^{2}+\mathbf{I}_{a}^{2}}}\,.
\end{equation}
%and the treatment given in Sec. \ref{BPS-Case-1}
%should be changed accordingly.

\paragraph{Single-Center Solution and MS Wall}

Clearly, in this case the limit $\mathbf{I}_{2}\rightarrow 0^{+}$ of Eqs. (%
\ref{I}) and (\ref{S-ineq-BPS}), and related comments, hold, as well.

Within the general condition (\ref{Q2-BPS-small-2}) on $\mathcal{Q}_{2}$
within $\mathbb{CP}^{1}$ (namely, by assuming the existence of a stability
region for the two-center configuration ``large'' BPS $+$ ``small'' BPS (\ref
{2-BPS})), the existence of a BPS MS wall $MS_{BPS}$ (\textit{cfr.} the
limit $\mathbf{I}_{2}\rightarrow 0^{+}$ of (\ref{q0Q0>0})) implies the
existence of the ``large'' BPS single-center solution with charge $\mathcal{Q%
}_{1}+\mathcal{Q}_{2}$, with entropy strictly larger than the entropy of the
two-center solution, as given by the limit $\mathbf{I}_{2}\rightarrow 0^{+}$
of Eq. (\ref{PA-2}).

Since one of the two centers is ``small'', this case is similar to the one
treated \textit{e.g.} in Sec. 5 of \cite{BD-1}, with the important
difference that for the $\mathbb{CP}^{1}$ model under consideration the
corresponding existing single-center solution is necessarily BPS with
entropy larger than the corresponding two-center solution (see the
discussion in Sec. \ref{Intro}, as well as the comment below Eq. (\ref
{Quart-in})).

\subsubsection{\label{BPS-Case-3}Case \textbf{3}}

This case cannot be consistently obtained by performing the $\mathbf{I}%
_{1}\rightarrow 0^{+}$ limit of the treatment of case \textbf{2 }given in
Sec. \ref{BPS-Case-3}, due to the $1$-charge nature of the charge vector $%
\mathcal{Q}_{1}$ given by (\ref{Q1-BPS-large}).

On the other hand, it is immediate to realize that the most general charge
configuration supporting the two-center solution (\ref{3-BPS}) is
duality-related to
\begin{eqnarray}
\mathcal{Q}_{1} &\equiv &\left( 0,0,q_{0},q_{1}\right) \Rightarrow \left\{
\begin{array}{l}
\mathbf{I}_{1}=q_{0}^{2}-q_{1}^{2}\equiv 0\Leftrightarrow \left|
q_{0}\right| \equiv \left| q_{1}\right| ; \\
\nexists t_{H}\left( \mathcal{Q}_{1}\right) ;
\end{array}
\right.  \label{Q1-BPS-small} \\
&&  \notag \\
\mathcal{Q}_{2} &\equiv &\left( P^{0},0,0,Q_{1}\right) \Rightarrow \left\{
\begin{array}{l}
\mathbf{I}_{2}=\left( P^{0}\right) ^{2}-Q_{1}^{2}\equiv 0\Leftrightarrow
\left| P^{0}\right| \equiv \left| Q_{1}\right| ; \\
\nexists t_{H}\left( \mathcal{Q}_{2}\right) ,
\end{array}
\right.  \label{Q2-BPS-small}
\end{eqnarray}
which can thus be considered without any loss in generality. Indeed, for the
charge configuration (\ref{Q1-BPS-small})-(\ref{Q2-BPS-small}), besides $%
\mathbf{I}_{1}=\mathbf{I}_{2}=0$, it holds that:
\begin{equation}
\left( \mathcal{Q}_{1},\mathcal{Q}_{2}\right) :\left\{
\begin{array}{l}
\mathbf{I}_{s}=-q_{1}Q_{1}; \\
\mathbf{I}_{a}=-q_{0}P^{0}.
\end{array}
\right.
\end{equation}
Note that
\begin{equation}
\left\{
\begin{array}{l}
\mathbf{I}_{1}=0; \\
\mathbf{I}_{2}=0;
\end{array}
\right. \Rightarrow \mathbf{I}_{s}^{2}=\mathbf{I}_{a}^{2}.
\end{equation}

Within the configuration (\ref{Q1-BPS-small})-(\ref{Q2-BPS-small}), the real
and imaginary part of $Z_{1}\overline{Z_{2}}$ respectively read (recall (\ref
{Y-call}) and (\ref{X-call})):
\begin{eqnarray}
\text{Re}\left( Z_{1}\overline{Z_{2}}\right) &=&\frac{q_{1}Q_{1}}{2}\frac{%
\left( b^{2}+\frac{q_{0}}{q_{1}}b+a^{2}+\frac{P^{0}}{Q_{1}}a\right) }{\left(
1-b^{2}-a^{2}\right) };  \notag  \label{Re} \\
&& \\
\text{Im}\left( Z_{1}\overline{Z_{2}}\right) &=&-\frac{q_{0}P^{0}}{2}\frac{%
\left( 1+\frac{q_{1}}{q_{0}}b+\frac{Q_{1}}{P^{0}}a\right) }{\left(
1-b^{2}-a^{2}\right) }.  \label{Im}
\end{eqnarray}

The region of stability $\mathcal{S}_{BPS}$ defined in (\ref{S-BPS})
corresponds to the region of $\mathbb{CP}^{1}$ in which the inequality
\begin{equation}
\mathcal{S}_{BPS}:1\pm b\pm a<0  \label{inn-1}
\end{equation}
is satisfied. Note that in the second step we used $\mathbf{I}_{1}=\mathbf{I}%
_{2}=0$, and the two ``$\pm $'' are reciprocally independent, depending on
the signs of $q_{0}q_{1}$ and $P^{0}Q_{1}$, respectively. By solving (\ref
{inn-1}) in a consistent way with the metric constraint (\ref
{metric-domain-x-y}), one achieves the following manifestly $U\left(
1,1\right) $-invariant result:
\begin{eqnarray}
\mathcal{S}_{BPS} &:&\left\{
\begin{array}{l}
\mathbf{I}_{s}\mathbf{I}_{a}>0:\left\{
\begin{array}{l}
-\sqrt{1-b^{2}}<a<-\left( 1+b\right) ; \\
b\in \left( -1,0\right) ;
\end{array}
\right. \cup \left\{
\begin{array}{l}
1-b<a<\sqrt{1-b^{2}}; \\
b\in \left( 0,1\right) .
\end{array}
\right. \\
\\
\mathbf{I}_{s}\mathbf{I}_{a}<0:\left\{
\begin{array}{l}
1+b<a<\sqrt{1-b^{2}}; \\
b\in \left( -1,0\right) ;
\end{array}
\right. \cup \left\{
\begin{array}{l}
-\sqrt{1-b^{2}}<a<-\left( 1+b\right) ; \\
b\in \left( 0,1\right) ;
\end{array}
\right.
\end{array}
\right.  \notag \\
&&  \label{inn-3}
\end{eqnarray}

Then, one can study the existence of the BPS MS and AMS walls, defined by (%
\ref{MS-BPS})-(\ref{Y-call}). By solving the condition Im$\left( Z_{1}%
\overline{Z_{2}}\right) =0$ consistently with the metric constraint (\ref
{metric-domain-x-y}) yields to the following manifestly $U\left( 1,1\right) $%
-invariant result:
\begin{equation}
\left. \text{Re}\left( Z_{1}\overline{Z_{2}}\right) \right| _{\text{Im}%
\left( Z_{1}\overline{Z_{2}}\right) =0}=\frac{\mathbf{I}_{s}}{2}.
\label{ress}
\end{equation}
Therefore, one can formulate the conditions of existence of the BPS MS
\textit{or} AMS wall in the manifestly $U\left( 1,1\right) $-invariant
following way:
\begin{equation}
\mathbf{I}_{s}>0:\left\{
\begin{array}{l}
MS_{BPS}=\left\{
\begin{array}{l}
\mathbf{I}_{a}>0:\left\{
\begin{array}{l}
a=-1-b \\
b\in \left( -1,0\right) ;
\end{array}
\cup \left\{
\begin{array}{l}
a=1-b \\
b\in \left( 0,1\right) ;
\end{array}
\right. \right. \\
\\
\mathbf{I}_{a}<0:\left\{
\begin{array}{l}
a=1+b \\
b\in \left( -1,0\right) ;
\end{array}
\right. \cup \left\{
\begin{array}{l}
a=-1+b \\
b\in \left( 0,1\right) ;
\end{array}
\right.
\end{array}
\right. \\
\\
\nexists ~AMS_{BPS};
\end{array}
\right.  \label{Case-3-Is>0}
\end{equation}
\begin{equation}
\mathbf{I}_{s}<0:\left\{
\begin{array}{l}
\nexists ~MS_{BPS}; \\
\\
AMS_{BPS}=\left\{
\begin{array}{l}
\mathbf{I}_{a}>0:\left\{
\begin{array}{l}
a=1+b \\
b\in \left( -1,0\right) ;
\end{array}
\cup \left\{
\begin{array}{l}
a=-1+b \\
b\in \left( 0,1\right) ;
\end{array}
\right. \right. \\
\\
\mathbf{I}_{a}<0:\left\{
\begin{array}{l}
a=-1-b \\
b\in \left( -1,0\right) ;
\end{array}
\right. \cup \left\{
\begin{array}{l}
a=1-b \\
b\in \left( 0,1\right) .
\end{array}
\right.
\end{array}
\right.
\end{array}
\right.  \label{Case-3-Is<0}
\end{equation}

\paragraph{Single-Center Solution and MS Wall}

By recalling (\ref{I2(Q1+Q2)}), in this case it follows that
\begin{equation}
\mathbf{I}_{s}>0  \label{condddd}
\end{equation}
is also the general condition of existence of the ``large'' BPS
single-center solution with charge $\mathcal{Q}_{1}+\mathcal{Q}_{2}$. Thus,
the unique possibility is
\begin{equation}
\frac{1}{\pi}S_{\text{1-ctr,BPS}}\left( \mathcal{Q}_{1}+\mathcal{Q}%
_{2}\right) =\mathbf{I}_{s}>S_{\text{2-ctr,BPS}}\left( \mathcal{Q}_{1},%
\mathcal{Q}_{2}\right) =0.
\end{equation}

The results holding for cases \textbf{1} and \textbf{2} respectively treated
in\ Secs. \ref{BPS-Case-1} and \ref{BPS-Case-2} still hold for this case:
within the general conditions (\ref{Q1-BPS-small}) and (\ref{Q2-BPS-small})
on $\mathcal{Q}_{1}$ and $\mathcal{Q}_{2}$ within $\mathbb{CP}^{1}$ (namely,
by assuming the existence of a stability region for the two-center
configuration ``small'' BPS $+$ ``small'' BPS (\ref{3-BPS})), the condition
of existence of a BPS MS wall $MS_{BPS}$ (see (\ref{Case-3-Is>0})) matches
the condition (\ref{condddd}) of existence of the ``large'' BPS
single-center solution with charge $\mathcal{Q}_{1}+\mathcal{Q}_{2}$, with
entropy $\mathbf{I}_{s}>0$ strictly larger than the entropy of the
two-center solution. Indeed, the limit $\mathbf{I}_{1},\mathbf{I}%
_{2}\rightarrow 0^{+}$ of Eq. (\ref{PA-2}) trivially yields that the entropy
of the two-center solution vanishes.

This case is similar to the ones treated \textit{e.g.} in \cite
{GLS-2,Castro-Simon}, with the important difference that for the $\mathbb{CP}%
^{1}$ model under consideration the corresponding existing single-center
solution is necessarily BPS (see the discussion in Sec. \ref{Intro}).

\subsection{\label{Axion-Dilaton-nBPS}Non-BPS MS \textit{or} AMS Wall}

Within this Subsection, we assume
\begin{equation}
\left( \mathcal{Q}_{1},\mathcal{Q}_{2}\right) :\left\{
\begin{array}{l}
\mathbf{I}_{1}<0; \\
\\
\mathbf{I}_{2}<0,
\end{array}
\right.  \label{non-BPS-charges}
\end{equation}
as well as $\mathbb{CP}^{1}$ to be the \textit{spatially asymptotical}
scalar manifold (as in Sec. \ref{Axion-Dilaton-BPS}). Only one possibility
\textit{a priori} exists, namely:
\begin{equation}
\text{non-BPS~``large''}\rightarrow \text{non-BPS~``large''}+\text{%
non-BPS~``large''}.  \label{non-BPS}
\end{equation}

A crucial observations (not holding for the \textit{minimally coupled}
models with $n\geqslant 2$ complex scalars, treated in Sec. \ref{n>1}) is
that one can switch from $\mathcal{I}_{2}\left( \mathcal{Q}\right) >0$
(``large'' BPS BH states) to $\mathcal{I}_{2}\left( \mathcal{Q}\right) <0$
(``large'' non-BPS BH states) \textit{e.g.} by performing the following
simple transformation on the charge vector:
\begin{equation}
\mathcal{Q}\equiv \left( p^{0},p^{1},q_{0},q_{1}\right) ^{T}\rightarrow
\left( \pm p^{1},\pm p^{0},\pm q_{0},\pm q_{1}\right) ^{T},  \label{Q-transf}
\end{equation}
where all ``$\pm $'''s in the r.h.s. are reciprocally independent. The
relevant transformation for the following treatment is the one with all ``$+$%
'' or all ``$-$'' in the r.h.s. of (\ref{Q-transf}). Without any loss of
generality, we will consider the one with all ``$+$'''s:
\begin{equation}
\mathcal{Q}\equiv \left( p^{0},p^{1},q_{0},q_{1}\right) ^{T}\rightarrow
\left( p^{1},p^{0},q_{0},q_{1}\right) ^{T},  \label{Q-transf-++++}
\end{equation}
which is ultimately equivalent to the interchanging the $\mathcal{N}=2$
graviphoton with the Maxwell field of the \textit{minimally coupled} vector
multiplet.

By performing transformation (\ref{Q-transf-++++}) on both $\mathcal{Q}_{1}$
and $\mathcal{Q}_{2}$, the symplectic product $\left\langle \mathcal{Q}_{1},%
\mathcal{Q}_{2}\right\rangle $ gets unchanged, $t_{BPS}\rightarrow t_{nBPS}$%
, and\footnote{%
Note that (\ref{Z-mapping}) is consistent with the treatment given \textit{%
e.g.} in \cite{Gnecchi-2} (see for instance Eq. (5.5) therein).}
\begin{equation}
Z\rightarrow i\overline{D_{\widehat{t}}Z},  \label{Z-mapping}
\end{equation}
where $D_{\widehat{t}}Z$ is the \textit{``flat'' matter charge}:
\begin{equation}
D_{\widehat{t}}Z\equiv e_{\widehat{t}}^{t}D_{t}Z=i\left( 1-t\overline{t}%
\right) D_{t}Z.
\end{equation}

As a consequence, the known BPS formul\ae\ (\ref{BPS-distance}) and (\ref
{BPS-J}) \cite{D-1,BD-1} get mapped into their corresponding non-BPS
counterparts, namely (we use the notation $D_{\widehat{t}}Z_{\mathbf{a}%
}\equiv D_{\widehat{t}}Z\left( b,a;\mathcal{Q}_{\mathbf{a}}\right) $, $%
\mathbf{a}=1,2$ throughout):
\begin{eqnarray}
\left| \overrightarrow{x}_{1}-\overrightarrow{x}_{2}\right| &=&-\frac{%
\left\langle \mathcal{Q}_{1},\mathcal{Q}_{2}\right\rangle }{2}\frac{\left|
D_{\widehat{t}}Z_{1}+D_{\widehat{t}}Z_{2}\right| }{\text{Im}\left( D_{%
\widehat{t}}Z_{1}\overline{D_{\widehat{t}}Z_{2}}\right) }; \\
&&  \notag \\
\overrightarrow{J} &=&-\frac{\text{Im}\left( D_{\widehat{t}}Z_{1}\overline{%
D_{\widehat{t}}Z_{2}}\right) }{\left| D_{\widehat{t}}Z_{1}+D_{\widehat{t}%
}Z_{2}\right| }\left( \overrightarrow{x}_{1}-\overrightarrow{x}_{2}\right) .
\end{eqnarray}

By applying (\ref{Q-transf-++++}) to (\ref{MS-BPS}) and (\ref{AMS-BPS}),
also the definitions of non-BPS MS and AMS walls can thus be given (within $%
\mathbb{CP}^{1}$):
\begin{eqnarray}
MS_{nBPS} &\equiv &\left\{ b+ia:\left[
\begin{array}{l}
\text{Im}\left( D_{\widehat{t}}Z_{1}\overline{D_{\widehat{t}}Z_{2}}\right)
=0; \\
\\
\text{Re}\left( D_{\widehat{t}}Z_{1}\overline{D_{\widehat{t}}Z_{2}}\right)
>0;
\end{array}
\right. \right\} ;  \label{MS-nBPS} \\
&&  \notag \\
AMS_{nBPS} &\equiv &\left\{ b+ia:\left[
\begin{array}{l}
\text{Im}\left( D_{\widehat{t}}Z_{1}\overline{D_{\widehat{t}}Z_{2}}\right)
=0; \\
\\
\text{Re}\left( D_{\widehat{t}}Z_{1}\overline{D_{\widehat{t}}Z_{2}}\right)
<0;
\end{array}
\right. \right\} ,  \label{AMS-nBPS}
\end{eqnarray}
where
\begin{eqnarray}
2\left( 1-b^{2}-a^{2}\right) \text{Re}\left( D_{\widehat{t}}Z_{1}\overline{%
D_{\widehat{t}}Z_{2}}\right) &=&\left[
\begin{array}{l}
q_{1}Q_{1}+p^{1}P^{1} \\
+\left( q_{0}Q_{0}+p^{0}P^{0}\right) b^{2}+\left(
q_{1}Q_{0}+q_{0}Q_{1}-p^{1}P^{0}-p^{0}P^{1}\right) b \\
+\left( q_{0}Q_{0}+p^{0}P^{0}\right) a^{2}+\left(
q_{1}P^{0}+q_{0}P^{1}+p^{1}Q_{0}+p^{0}Q_{1}\right) a
\end{array}
\right] ;  \notag \\
&&  \label{Y-call-hat} \\
-2\left( 1-b^{2}-a^{2}\right) \text{Im}\left( D_{\widehat{t}}Z_{1}\overline{%
D_{\widehat{t}}Z_{2}}\right) &=&\left[
\begin{array}{l}
p^{1}Q_{1}-q_{1}P^{1} \\
+\left( q_{0}P^{0}-p^{0}Q_{0}\right) b^{2}+\left(
q_{1}P^{0}-q_{0}P^{1}+p^{1}Q_{0}-p^{0}Q_{1}\right) b \\
+\left( q_{0}P^{0}-p^{0}Q_{0}\right) a^{2}+\left(
-q_{1}Q_{0}+q_{0}Q_{1}+p^{1}P^{0}-p^{0}P^{1}\right) a
\end{array}
\right] .  \notag \\
&&  \label{X-call-hat}
\end{eqnarray}

Analogously, by applying (\ref{Q-transf-++++}) to (\ref{S-BPS}), the region
of stability $\mathcal{S}_{nBPS}\left( b,a;\mathcal{Q}_{1},\mathcal{Q}%
_{2}\right) $ of the two-center non-BPS solution can be defined as
\begin{equation}
\mathcal{S}_{nBPS}\equiv \left\{ b+ia\in \mathbb{CP}^{1}:\left\langle
\mathcal{Q}_{1},\mathcal{Q}_{2}\right\rangle \text{Im}\left( D_{\widehat{t}%
}Z_{1}\overline{D_{\widehat{t}}Z_{2}}\right) <0\right\} .  \label{S-nBPS}
\end{equation}

It is also worth observing that the ``large'' non-BPS single-center solution
with charge $\mathcal{Q}=\mathcal{Q}_{1}+\mathcal{Q}_{2}$ would exist
\textit{iff} (recall (\ref{I2(Q1+Q2)}))
\begin{equation}
\mathcal{I}_{2}\left( \mathcal{Q}_{1}+\mathcal{Q}_{2}\right)
<0\Leftrightarrow 2\mathbf{I}_{s}<-\left( \mathbf{I}_{1}+\mathbf{I}%
_{2}\right) ,  \label{nBPS-single-cond}
\end{equation}
where the condition (\ref{non-BPS-charges}) must be taken into account.

\subsubsection{\label{Non-BPS}Analysis}

The most general charge configuration supporting the two-center solution (%
\ref{non-BPS}) is duality-related to
\begin{eqnarray}
\mathcal{Q}_{1} &\equiv &\left( 0,0,0,q_{1}\right) \Rightarrow \left\{
\begin{array}{l}
\mathbf{I}_{1}=-q_{1}^{2}<0; \\
\\
t_{H,nBPS}\left( \mathcal{Q}_{1}\right) =0;
\end{array}
\right.  \label{Q1-nBPS} \\
&&  \notag \\
\mathcal{Q}_{2} &\equiv &\left( P^{0},P^{1},0,Q_{1}\right) \Rightarrow
\left\{
\begin{array}{l}
\mathbf{I}_{2}=\left( P^{0}\right) ^{2}-\left( P^{1}\right) ^{2}-Q_{1}^{2}<0;
\\
\\
t_{H,nBPS}\left( \mathcal{Q}_{2}\right) =-i\frac{P^{0}}{\left(
Q_{1}-iP^{1}\right) },
\end{array}
\right.  \label{Q2-nBPS-cond}
\end{eqnarray}
which can thus be considered without any loss in generality. Indeed, for the
charge configuration (\ref{Q1-nBPS})-(\ref{Q2-nBPS-cond}), the four
quadratic $U\left( 1,1\right) $-invariants (\ref{1})-(\ref{4}) are all
generally non-coinciding and non-vanishing:
\begin{equation}
\left( \mathcal{Q}_{1},\mathcal{Q}_{2}\right) :\left\{
\begin{array}{l}
\mathbf{I}_{1}=-q_{1}^{2}; \\
\mathbf{I}_{2}=\left( P^{0}\right) ^{2}-\left( P^{1}\right) ^{2}-Q_{1}^{2}<0;
\\
\mathbf{I}_{s}=-q_{1}Q_{1}; \\
\mathbf{I}_{a}=-q_{1}P^{1}.
\end{array}
\right.  \label{n=1-nBPS-invs}
\end{equation}
A manifestly $U\left( 1,1\right) $-invariant characterization of the four
non-vanishing charges of the general non-BPS two-center configuration (\ref
{Q1-nBPS})-(\ref{Q2-nBPS-cond}) reads as follows:
\begin{eqnarray}
q_{1}^{2} &=&-\mathbf{I}_{1};  \label{q1-inv-nBPS} \\
\left( P^{0}\right) ^{2} &=&\frac{\left( \mathbf{I}_{1}\mathbf{I}_{2}-%
\mathbf{I}_{s}^{2}-\mathbf{I}_{a}^{2}\right) }{\mathbf{I}_{1}};
\label{P0-inv-nBPS} \\
\left( P^{1}\right) ^{2} &=&-\frac{\mathbf{I}_{a}^{2}}{\mathbf{I}_{1}};
\label{P1-inv-nBPS} \\
Q_{1}^{2} &=&-\frac{\mathbf{I}_{s}^{2}}{\mathbf{I}_{1}},  \label{Q1-inv-nBPS}
\end{eqnarray}
where
\begin{equation}
\left.
\begin{array}{r}
\mathbf{I}_{1}<0 \\
\mathbf{I}_{2}<0
\end{array}
\right\} \Rightarrow \mathbf{I}_{1}\mathbf{I}_{2}-\mathbf{I}_{s}^{2}-\mathbf{%
I}_{a}^{2}<0.
\end{equation}

Note that the configuration (\ref{Q1-nBPS})-(\ref{Q2-nBPS-cond}) (and in
general all the treatment of non-BPS case given below) can be obtained from (%
\ref{Q1-BPS-large})-(\ref{Q2-BPS-large-cond}) (and in general all the
treatment of BPS case given in Sec. \ref{BPS-Case-1}) by performing the
transformation (\ref{Q-transf-++++}) on both $\mathcal{Q}_{1}$ and $\mathcal{%
Q}_{2}$.

Within the configuration (\ref{Q1-nBPS})-(\ref{Q2-nBPS-cond}), the real and
imaginary part of $D_{\widehat{t}}Z_{1}\overline{D_{\widehat{t}}Z_{2}}$
respectively read (recall (\ref{Y-call-hat}) and (\ref{X-call-hat})):
\begin{eqnarray}
\text{Re}\left( D_{\widehat{t}}Z_{1}\overline{D_{\widehat{t}}Z_{2}}\right)
&=&\frac{q_{1}\left( Q_{1}+P^{0}a\right) }{2\left( 1-b^{2}-a^{2}\right) };
\label{Y-call-hat-simpl} \\
\text{Im}\left( D_{\widehat{t}}Z_{1}\overline{D_{\widehat{t}}Z_{2}}\right)
&=&\frac{q_{1}\left( P^{1}-P^{0}b\right) }{2\left( 1-b^{2}-a^{2}\right) }.
\label{X-call-hat-simpl}
\end{eqnarray}

Let us start by computing the region of stability $\mathcal{S}_{nBPS}$
defined in (\ref{S-nBPS}):
\begin{equation}
\mathcal{S}_{nBPS}:\frac{P^{0}}{P^{1}}b<1\Leftrightarrow \left\{
\begin{array}{l}
P^{0}P^{1}>0:-\sqrt{1-a^{2}}<b<\frac{P^{1}}{P^{0}}<\sqrt{1-a^{2}}; \\
\\
P^{0}P^{1}<0:-\sqrt{1-a^{2}}<\frac{P^{1}}{P^{0}}<b<\sqrt{1-a^{2}}.
\end{array}
\right.  \label{S-nBPS-gen}
\end{equation}
Note that $a$ enters (\ref{S-nBPS-gen}) only through the constraint to
belong to the domain of definition of the metric of the scalar manifold,
defined by (\ref{metric-domain-x-y}), which in this case implies
\begin{equation}
\left| \frac{P^{1}}{P^{0}}\right| <1\Leftrightarrow \frac{\left| \mathbf{I}%
_{a}\right| }{\sqrt{\mathbf{I}_{s}^{2}+\mathbf{I}_{a}^{2}-\mathbf{I}_{1}%
\mathbf{I}_{2}}}<1.  \label{impl-nBPS}
\end{equation}

By using (\ref{P0-inv-nBPS})-(\ref{P1-inv-nBPS}), the region of stability $%
\mathcal{S}_{nBPS}$ (\ref{S-nBPS-gen})-(\ref{impl-nBPS}) can be re-expressed
as follows:
\begin{equation}
\mathcal{S}_{nBPS}:\left\{
\begin{array}{l}
\pm \frac{\sqrt{\mathbf{I}_{s}^{2}+\mathbf{I}_{a}^{2}-\mathbf{I}_{1}\mathbf{I%
}_{2}}}{\left| \mathbf{I}_{a}\right| }b<1; \\
\\
b^{2}+a^{2}<1,
\end{array}
\right.  \label{S-nBPS-gen-inv}
\end{equation}
which is remarkably symmetric with respect to its BPS counterpart given by (%
\ref{S-BPS-gen-inv}).

Then, one can study the existence of the non-BPS\ MS and AMS walls, defined
by (\ref{MS-nBPS})-(\ref{Y-call-hat}). Within the condition (\ref{impl-nBPS}%
), it is convenient to introduce
\begin{eqnarray}
a_{nBPS} &\equiv &\sqrt{\frac{\mathbf{I}_{1}\mathbf{I}_{2}-\mathbf{I}_{s}^{2}%
}{\mathbf{I}_{1}\mathbf{I}_{2}-\mathbf{I}_{s}^{2}-\mathbf{I}_{a}^{2}}}>0;
\label{def-a_nBPS} \\
&&  \notag \\
&&  \notag \\
\mathcal{B} &\equiv &\left\{ b,a\in \mathbb{CP}^{1}:\left[
\begin{array}{l}
b=\pm \frac{\left| \mathbf{I}_{a}\right| }{\sqrt{\mathbf{I}_{s}^{2}+\mathbf{I%
}_{a}^{2}-\mathbf{I}_{1}\mathbf{I}_{2}}}; \\
\\
-\left| \frac{Q_{1}}{P^{0}}\right| <-a_{nBPS}<a<a_{nBPS}<\left| \frac{Q_{1}}{%
P^{0}}\right| .
\end{array}
\right. \right\} =\left. \mathcal{A}\right| _{P^{0}\leftrightarrow P^{1}};
\label{B-call} \\
&&  \notag \\
\left| \frac{Q_{1}}{P^{0}}\right| &=&\frac{\left| \mathbf{I}_{s}\right| }{%
\sqrt{\mathbf{I}_{s}^{2}+\mathbf{I}_{a}^{2}-\mathbf{I}_{1}\mathbf{I}_{2}}}.
\end{eqnarray}
Then, through some straightforward computations (detailed in App. B), one
obtains that within the two-center charge configuration (\ref{Q1-nBPS})-(\ref
{Q2-nBPS-cond}) the existence of non-BPS MS or AMS walls depends on the sign
of $\mathbf{I}_{s}$:
\begin{eqnarray}
\mathbf{I}_{s} &<&0:\left\{
\begin{array}{l}
MS_{nBPS}=\mathcal{B}; \\
\\
\nexists AMS_{nBPS};
\end{array}
\right.  \label{q1Q1>0} \\
&&  \notag \\
\mathbf{I}_{s} &>&0:\left\{
\begin{array}{l}
\nexists MS_{nBPS}; \\
\\
AMS_{nBPS}=\mathcal{B},
\end{array}
\right.  \label{q1Q1<0}
\end{eqnarray}
It is interesting to compare Eqs. (\ref{def-a_nBPS})-(\ref{q1Q1<0}) with
their BPS counterparts, respectively given by (\ref{def-a_BPS})-(\ref{q0Q0<0}%
).

\paragraph{Single-Center Solution and MS Wall}

\begin{equation}
\mathbf{I}_{1}+\mathbf{I}_{2}+2\mathbf{I}_{s}<0  \label{VI}
\end{equation}
is the the general condition (\ref{nBPS-single-cond}) of existence of the
``large'' non-BPS single-center solution with charge $\mathcal{Q}_{1}+%
\mathcal{Q}_{2}$. Notice that
\begin{equation}
S_{\text{1-ctr,nBPS}}\left( \mathcal{Q}_{1}+\mathcal{Q}_{2}\right)
\gtreqless S_{\text{2-ctr,nBPS}}\left( \mathcal{Q}_{1},\mathcal{Q}%
_{2}\right) \Leftrightarrow \left\{
\begin{array}{l}
\mathbf{I}_{s}<0; \\
\mathbf{I}_{s}=0; \\
0<\mathbf{I}_{s}<-\frac{1}{2}\left( \mathbf{I}_{1}+\mathbf{I}_{2}\right) .
\end{array}
\right.  \label{S-ineq-nBPS}
\end{equation}

\textit{Mutatis mutandis}, the story goes as in the BPS case treated in Sec.
\ref{BPS-Case-1}. Indeed, as anticipated in\ Sec. \ref{Intro}, within the
general conditions (\ref{Q2-nBPS-cond}) and (\ref{impl-nBPS}) on $\mathcal{Q}%
_{2}$ (corresponding to assuming the existence of a stability region for the
two-center configuration ``large'' non-BPS $+$ ``large'' non-BPS (\ref
{non-BPS})), the existence of a non-BPS MS wall $MS_{nBPS}$ (see (\ref
{q1Q1>0})) implies the existence of the ``large'' non-BPS single-center
solution with charge $\mathcal{Q}_{1}+\mathcal{Q}_{2}$, with entropy
strictly larger than the entropy of the two-center solution, as given by Eq.
(\ref{PA-2-2}).\medskip

Note that the analysis of Sec. \ref{Axion-Dilaton-nBPS} provides the an
example worked out in full generality of non-BPS two-center BH solution with
constrained positions of the centers (\textit{i.e.} mutually non-local
charges $\mathcal{Q}_{1}$ and $\mathcal{Q}_{2}$). In fact, it is worth
pointing out that this case does not belong to the class of non-BPS
multi-center solutions studied \textit{e.g.} in \cite{G-1} and \cite{GP-1},
nor to the $\mathcal{I}_{4}<0$ two-center solution of \cite{Bena:2009ev}.

This is also due to the fact that the $t^{2}$ model is the unique known
model in which the non-BPS fake superpotential is the absolute value of a
complex quantity linear in the charges, namely \cite{ADOT-1,Gnecchi-1}
\begin{equation}
W_{nBPS}=\left| D_{\widehat{t}}Z\right| .  \label{n=1-exc}
\end{equation}
This remarkably form of $W_{nBPS}$ allowed for an especially simple
treatment of non-BPS two-center solution in full generality in Sec. \ref
{Axion-Dilaton-nBPS}.

\section{\label{n>1}Many Moduli}

We now turn to consider the $\mathcal{N}=2$, $d=4$ supergravity models with $%
n\geqslant 2$ Abelian vector multiplets \textit{minimally coupled} to the
gravity multiplet \cite{Luciani}. The metric of the scalar manifold can be
computed to read \cite{Luciani,Gnecchi-1} (Einstein summation convention on
repeated indices is used, and $i=1,...,n$, throughout):
\begin{eqnarray}
g_{i\overline{j}} &\equiv &\partial _{i}\overline{\partial }_{\overline{j}}%
\mathcal{K}=\frac{\left( 1-\left| t^{l}\right| ^{2}\right) \delta _{i%
\overline{j}}+\overline{t}^{\overline{i}}t^{j}}{\left( 1-\left| t^{k}\right|
^{2}\right) ^{2}}=2e^{\mathcal{K}}\delta _{i\overline{j}}+4e^{2\mathcal{K}}%
\overline{t}^{\overline{i}}t^{j}; \\
g^{i\overline{j}} &=&\left( 1-\left| t^{k}\right| ^{2}\right) \left( \delta
^{i\overline{j}}-t^{i}\overline{t}^{\overline{j}}\right) =\frac{1}{2}e^{-%
\mathcal{K}}\left( \delta ^{i\overline{j}}-t^{i}\overline{t}^{\overline{j}%
}\right) ; \\
g_{i\overline{j}}g^{i\overline{k}} &=&\delta _{\overline{j}}^{\overline{k}}.
\end{eqnarray}

The domain of definition of the K\"{a}hler potential $\mathcal{K}$ and of
the metric $g_{i\overline{j}}$ is the interior of the $2n$-hypersphere of
unitary radius centered in the origin:
\begin{equation}
\sum_{i=1}^{n}\left| t^{i}\right| ^{2}<1.  \label{metric-domain-n>1}
\end{equation}

The expressions of the central charge and of the \textit{matter charges} are
given by Eqs. (\ref{Z-CP^n})-(\ref{DiZ-CP^n}), whereas the BPS and non-BPS
attractor values of scalar fields, respectively read as follows \cite
{Gnecchi-1}:
\begin{eqnarray}
t_{BPS}^{i} &=&-\frac{\left( q_{i}+ip^{i}\right) }{q_{0}-ip^{0}},~\forall i;
\label{t^i_BPS} \\
t_{nBPS}^{1} &=&-\frac{\left( q_{a}-ip^{a}\right) t_{nBPS}^{a}}{q_{1}-ip^{1}}%
-\frac{q_{0}+ip^{0}}{q_{1}-ip^{1}}.  \label{t-non-BPS-n>1}
\end{eqnarray}
Without any loss of generality (up to re-labelling), in (\ref{t-non-BPS-n>1}%
) the complex scalar field $t^{1}$ is stabilized in terms of the non-BPS
values $t_{nBPS}^{a}$ of the remaining $n-1$ scalars. Notice that $%
t_{nBPS}^{a}$ are not fixed by the Attractor Mechanism; indeed, such scalars
are known to coordinatise the \textit{``moduli space''} of non-BPS attractor
solutions in \textit{minimally coupled} sequence, which is nothing but $%
\mathbb{CP}^{n-1}$ \cite{Ferrara-Marrani-2}
\begin{equation}
\left\{ t_{nBPS}^{a}\right\} _{a=2,...,n}\in \mathcal{M}_{nBPS}=\mathbb{CP}%
^{n-1}.  \label{CP^(n-1)}
\end{equation}
As evident from the treatment of Sec. \ref{n=1}, in the $1$-modulus case
there is no non-BPS $Z_{H}=0$ \textit{``moduli space''} at all.

It is worth remarking that the number of quadratic $U\left( 1,n\right) $%
-invariants does \textit{not} depend on the number $n$ of \textit{minimally
coupled} vector multiplets, and it is then always equal to four. Thus, the $%
n\geqslant 2$ generalization of the most general charge configuration\ (\ref
{Q1-BPS-large})-(\ref{Q2-BPS-large-cond}) supporting the two-center
``large'' BPS $+$ ``large'' BPS BH solution (\ref{1-BPS}) is duality-related
to ($a=2,...,n$ throughout)
\begin{eqnarray}
\mathcal{Q}_{1} &\equiv &\left( 0,p^{i}=0,q_{0},q_{i}=0\right) \Rightarrow
\mathbf{I}_{1}=q_{0}^{2}>0;  \label{Q1-BPS-large-n} \\
\mathcal{Q}_{2} &\equiv &\left( P^{0},P^{1},P^{a}=0,Q_{0},Q_{i}=0\right)
\Rightarrow \mathbf{I}_{2}=\left( P^{0}\right) ^{2}+Q_{0}^{2}-\left(
P^{1}\right) ^{2}>0,  \label{Q2-BPS-large-cond-n}
\end{eqnarray}
implying
\begin{equation}
\left\{
\begin{array}{l}
t_{BPS}^{i}\left( \mathcal{Q}_{1}\right) =0,~\forall i; \\
\\
t_{BPS}^{1}\left( \mathcal{Q}_{2}\right) =-\frac{iP^{1}}{\left(
Q_{0}-iP^{0}\right) }; \\
\\
t_{BPS}^{a}\left( \mathcal{Q}_{2}\right) =0,~\forall a.
\end{array}
\right.
\end{equation}
On the same respect, the $n\geqslant 2$ generalization of the most general
charge configuration (\ref{Q1-nBPS})-(\ref{Q2-nBPS-cond}) supporting the
two-center non-BPS BH solution (\ref{non-BPS})\ is duality-related to
\begin{eqnarray}
\mathcal{Q}_{1} &\equiv &\left( 0,p^{i}=0,0,q_{1},q_{a}=0\right) \Rightarrow
\mathbf{I}_{1}=-q_{1}^{2}<0;  \label{Q1-nBPS-n} \\
\mathcal{Q}_{2} &\equiv &\left( P^{0},P^{1},P^{a}=0,0,Q_{1},Q_{a}=0\right)
\Rightarrow \mathbf{I}_{1}=\left( P^{0}\right) ^{2}-\left( P^{1}\right)
^{2}-Q_{1}^{2}<0,  \label{Q2-nBPS-cond-n}
\end{eqnarray}
implying (recall (\ref{CP^(n-1)}))
\begin{equation}
\left\{
\begin{array}{l}
t_{nBPS}^{1}\left( \mathcal{Q}_{1}\right) =0; \\
\\
t_{nBPS}^{1}\left( \mathcal{Q}_{2}\right) =i\frac{\left(
P^{a}t_{nBPS}^{a}-P^{0}\right) }{\left( Q_{1}-iP^{1}\right) }; \\
\\
t_{nBPS}^{a}\in \mathbb{CP}^{n-1},~\forall a.
\end{array}
\right.
\end{equation}

Thus, both (\ref{Q1-BPS-large-n})-(\ref{Q2-BPS-large-cond-n}) and (\ref
{Q1-nBPS-n})-(\ref{Q2-nBPS-cond-n}) can be considered without any loss in
generality. As a consequence, the treatment of BPS MS/AMS in the case $%
n\geqslant 2$ (see Sec. \ref{n>1-BPS} further below) is very similar to the
treatment done in the case $n=1$ in Sec. \ref{Axion-Dilaton-BPS}, the main
difference consisting in the change of the metric constraint, which is now
given by (\ref{metric-domain-n>1}).

On the other hand, the treatment of non-BPS MS/AMS walls in the case $%
n\geqslant 2$ (see Sec. \ref{n>1-nBPS} further below) is different from the
treatment done in the case $n=1$ in Sec. \ref{Axion-Dilaton-BPS}. Indeed,
for $n\geqslant 2$ the accidental $n=1$ symmetry between the $\mathcal{N}=2$
central charge $Z$ and the \textit{``flat'' matter charge}\footnote{%
Recall Eq. (\ref{Z-mapping}). This can be interpreted as exchange of the two
skew-eigenvalues of the central charge matrix in the $\mathcal{N}=4$
supersymmetry uplift; see \textit{e.g.} Eq. (5.5) of \cite{Gnecchi-2}, as
well as \cite{FHM-rev} and Refs. therein.} $i\overline{D_{\widehat{t}}Z}$ is
spoiled. %Actually, the $n=1$ \textit{minimally coupled} model treated in
%Sec. \ref{n=1} is exceptional, in the sense that, to the best of our
%knowledge, it is the unique model in which the non-BPS superpotential $%
%W_{nBPS}$ is the absolute value of a complex quantity linear in charges, as
%given by Eq. (\ref{n=1-exc}) further below \cite{ADOT-1,Gnecchi-1}.

We will briefly consider the treatment of non-BPS two-center configuration
in the case $n\geqslant 2$, based on the results of \cite{MS-FM-1}, in Sec.
\ref{n>1-nBPS}.

\subsection{\label{n>1-BPS}BPS MS \textit{or} AMS\ Wall}

Within this Subsection, we assume $\mathcal{Q}_{1}$ and $\mathcal{Q}_{2}$ to
satisfy (\ref{BPS-charges}), as well as $\mathbb{CP}^{n}$ to be the \textit{%
spatially asymptotical} scalar manifold.

The \textit{a priori} possible BPS ``large'' two-center configurations are
given by (\ref{1-BPS})-(\ref{3-BPS}), with the BPS MS and AMS\ walls defined
by (\ref{MS-BPS}) and (\ref{AMS-BPS}) (clearly, with $\mathbb{CP}^{1}$
replaced by $\mathbb{CP}^{n}$). The $n\geqslant 2$ generalizations of the
explicit $n=1$ expressions (\ref{Y-call}) and (\ref{X-call}) are cumbersome
and, within the choice of charges (\ref{Q1-BPS-large-n})-(\ref
{Q2-BPS-large-cond-n}), useless; thus, we will refrain from reporting them
here.

The BPS stability region, the distance between the centers $1$ and $2$, and
the corresponding configurational angular momentum are still given by the
formul\ae\ (\ref{S-BPS})-(\ref{BPS-J}). Moreover, the condition of existence
of the ``large'' BPS single-center solution with charge $\mathcal{Q}=%
\mathcal{Q}_{1}+\mathcal{Q}_{2}$ is given by (\ref{large-BPS-single-cond}).

\subsubsection{\label{BPS-case-1-n>1}Case \textbf{1}}

Without any loss of generality, we consider the two-charge configuration (%
\ref{Q1-BPS-large-n})-(\ref{Q2-BPS-large-cond-n}). Within such a
configuration, the four quadratic $U\left( 1,n\right) $-invariants (\ref{1}%
)-(\ref{4}) are all generally non-coinciding and non-vanishing, and they do
match the expressions (\ref{n=1-BPS-invs}) holding for the case $n=1$
itself. Consequently, a manifestly $U\left( 1,n\right) $-invariant
characterization of the four non-vanishing charges of the general BPS
two-center configuration (\ref{Q1-BPS-large-n})-(\ref{Q2-BPS-large-cond-n})
is the very same as the $n=1$ one given by (\ref{q0-inv})-(\ref{PA-1}).

Within the general configuration (\ref{Q1-BPS-large-n})-(\ref
{Q2-BPS-large-cond-n}), one obtains that
\begin{eqnarray}
Z_{1} &=&\frac{q_{0}}{\sqrt{2}\sqrt{1-a^{2}-b^{2}-\left| t^{a}\right| ^{2}}};
\\
Z_{2} &=&\frac{\left[ Q_{0}+iP^{0}-iP^{1}\left( b+ia\right) \right] }{\sqrt{2%
}\sqrt{1-a^{2}-b^{2}-\left| t^{a}\right| ^{2}}},
\end{eqnarray}
which are thus coinciding, up to the different K\"{a}hler overall factor,
with their $n=1$ counterparts. As a consequence, it holds that the real and
imaginary part of $Z_{1}\overline{Z_{2}}$ respectively read:
\begin{eqnarray}
\text{Re}\left( Z_{1}\overline{Z}_{2}\right) &=&\frac{q_{0}\left(
Q_{0}+P^{1}a\right) }{2\left( 1-a^{2}-b^{2}-\left| t^{a}\right| ^{2}\right) }%
;  \label{Y-call-simpl-n} \\
\text{Im}\left( Z_{1}\overline{Z}_{2}\right) &=&\frac{q_{0}\left(
-P^{0}+P^{1}b\right) }{2\left( 1-a^{2}-b^{2}-\left| t^{a}\right| ^{2}\right)
},  \label{X-call-simpl-n}
\end{eqnarray}
which still match, up to the different K\"{a}hler overall factor, their $n=1$
counterparts, respectively given by (\ref{Y-call-simpl}) and (\ref
{X-call-simpl}).

By recalling (\ref{S-BPS-gen}), the region of stability $\mathcal{S}%
_{BPS}\left( \mathcal{Q}_{1},\mathcal{Q}_{2}\right) $ defined in (\ref{S-BPS}%
) can be easily computed to be
\begin{equation}
\mathcal{S}_{BPS}:\frac{P^{1}}{P^{0}}b>1\Leftrightarrow \left\{
\begin{array}{l}
P^{0}P^{1}>0:\frac{P^{0}}{P^{1}}<b<\sqrt{1-a^{2}-\left| t^{a}\right| ^{2}};
\\
\\
P^{0}P^{1}<0:-\sqrt{1-a^{2}-\left| t^{a}\right| ^{2}}<b<\frac{P^{0}}{P^{1}}.
\end{array}
\right.  \label{S-BPS-gen-n}
\end{equation}
Note that $a$ and the remaining $n-1$ complex fields $t^{a}$'s enter Eq. (%
\ref{S-BPS-gen-n}) only through the constrain to belong to the domain of
definition of the metric of the scalar manifold, defined by (\ref
{metric-domain-n>1}). By using (\ref{P0-inv})-(\ref{P1-inv}), the region of
stability $\mathcal{S}_{BPS}$ (\ref{S-BPS-gen-n}) can thus be re-expressed
through the $n\geqslant 2$ generalization of Eq. (\ref{S-BPS-gen-inv}):
\begin{equation}
\mathcal{S}_{BPS}:\left\{
\begin{array}{l}
\pm \frac{\sqrt{\mathbf{I}_{s}^{2}+\mathbf{I}_{a}^{2}-\mathbf{I}_{1}\mathbf{I%
}_{2}}}{\left| \mathbf{I}_{a}\right| }b>1; \\
\\
b^{2}+a^{2}+\left| t^{a}\right| ^{2}<1.
\end{array}
\right.  \label{S-BPS-gen-inv-n}
\end{equation}

Then, one can study the existence of the BPS\ MS and AMS walls, which are
defined by (\ref{MS-BPS}) and (\ref{AMS-BPS}) (with $\mathbb{CP}^{1}$
replaced by $\mathbb{CP}^{n}$).

By assuming the condition ($n\geqslant 2$ generalization of the (\ref{impl}%
))
\begin{equation}
\left( \mathbf{I}_{s}^{2}+\mathbf{I}_{a}^{2}-\mathbf{I}_{1}\mathbf{I}%
_{2}\right) \left( 1-\left| t^{a}\right| ^{2}\right) -\mathbf{I}_{a}^{2}>0,
\label{stricter-than-impl}
\end{equation}
one can recall (\ref{def-a_BPS}) and define
\begin{equation}
\mathcal{A}_{n}\equiv \left\{ \left\{ t^{i}\right\} _{i=1,...,n}\in \mathbb{%
CP}^{n}:\left[
\begin{array}{l}
b=\pm \frac{\left| \mathbf{I}_{a}\right| }{\sqrt{\mathbf{I}_{s}^{2}+\mathbf{I%
}_{a}^{2}-\mathbf{I}_{1}\mathbf{I}_{2}}}; \\
\\
-\sqrt{a_{BPS}^{2}-\left| t^{a}\right| ^{2}}<a<\sqrt{a_{BPS}^{2}-\left|
t^{a}\right| ^{2}}.
\end{array}
\right. \right\} ,  \label{def-A-call-n}
\end{equation}
which is the $n\geqslant 2$ generalization of the region $\mathcal{A}%
\subsetneq \mathbb{CP}^{1}$ defined in (\ref{def-A-call}). Thus, through
some straightforward computations (the $n\geqslant 2$ analogues of the ones
detailed in App. A), one obtains that within the two-center charge
configuration (\ref{Q1-BPS-large-n})-(\ref{Q2-BPS-large-cond-n}) the
existence of BPS MS or AMS walls depends on the sign of $\mathbf{I}_{s}$:
\begin{eqnarray}
\mathbf{I}_{s} &>&0:\left\{
\begin{array}{l}
MS_{BPS}=\mathcal{A}_{n}; \\
\\
\nexists AMS_{BPS};
\end{array}
\right.  \label{q0Q0>0-n} \\
&&  \notag \\
\mathbf{I}_{s} &<&0:\left\{
\begin{array}{l}
\nexists MS_{BPS}; \\
\\
AMS_{BPS}=\mathcal{A}_{n}.
\end{array}
\right.  \label{q0Q0<0-n}
\end{eqnarray}

\paragraph{Single-Center Solution and MS Wall}

Let us also remark that, within the configuration (\ref{Q1-BPS-large-n})-(%
\ref{Q2-BPS-large-cond-n}), Eq. (\ref{S-ineq-BPS}) keeps holding true.

Within the general conditions (\ref{Q2-BPS-large-cond-n}) and (\ref
{stricter-than-impl}) on $\mathcal{Q}_{2}$ (corresponding to assuming the
existence of a stability region for the two-center configuration ``large''
BPS $+$ ``large'' BPS (\ref{1-BPS}) in the case $n\geqslant 2$), the
existence of a BPS MS wall $MS_{BPS}$ (see (\ref{q0Q0>0-n})) implies the
existence of the ``large'' BPS single-center solution with charge $\mathcal{Q%
}_{1}+\mathcal{Q}_{2}$, with entropy strictly larger than the entropy of the
two-center solution, as given by Eq. (\ref{PA-2}).

\subsection{\label{n>1-nBPS}Non-BPS}

As mentioned above, for the study of two-center non-BPS solutions in
presence of $n\geqslant 2$ Abelian vector multiplets coupled to $\mathcal{N}%
=2 $, $d=4$ supergravity multiplet, a different approach with respect to the
case $n=1$ (treated in Sec. \ref{Axion-Dilaton-nBPS}) must be adopted.

This approach relies on the general formul\ae\ of the MS wall, AMS wall,
distance between centers $1$ and $2$, and configurational angular momentum
for two-center non-BPS ``large'' $+$ non-BPS ``large'' solutions in \textit{%
minimally coupled} $\mathcal{N}=2$, $d=4$ supergravity. By using the
notation
\begin{equation}
W\left( \left\{ t^{i},\overline{t}^{\overline{i}}\right\} _{i=1,...,n},%
\mathcal{Q}_{\mathbf{a}}\right) \equiv W_{\mathbf{a}},~\mathbf{a}=1,2,
\end{equation}
such formul\ae\ respectively read \cite{MS-FM-1}:
\begin{eqnarray}
MS_{nBPS} &:&W_{1+2}=W_{1}+W_{2};  \label{MS-nBPS-n} \\
&&  \notag \\
AMS_{nBPS} &:&W_{1+2}=\left| W_{1}-W_{2}\right| ;  \label{AMS-nBPS-n} \\
&&  \notag \\
\left| \overrightarrow{x_{1}}-\overrightarrow{x_{2}}\right| &=&\pm \frac{%
\left\langle \mathcal{Q}_{1},\mathcal{Q}_{2}\right\rangle W_{1+2}}{\sqrt{%
4W_{1}^{2}W_{2}^{2}-\left( W_{1+2}^{2}-W_{1}^{2}-W_{2}^{2}\right) ^{2}}};
\label{distance-nBPS-n} \\
&&  \notag \\
\overrightarrow{J} &=&\frac{\left\langle \mathcal{Q}_{1},\mathcal{Q}%
_{2}\right\rangle }{2}\frac{\left( \overrightarrow{x}_{1}-\overrightarrow{x}%
_{2}\right) }{\left| \overrightarrow{x}_{1}-\overrightarrow{x}_{2}\right| }%
=\pm \frac{\left( \overrightarrow{x}_{1}-\overrightarrow{x}_{2}\right) }{2}%
\frac{\sqrt{4W_{1}^{2}W_{2}^{2}-\left(
W_{1+2}^{2}-W_{1}^{2}-W_{2}^{2}\right) ^{2}}}{W_{1+2}},  \notag \\
&&  \label{J-nBPS-n}
\end{eqnarray}
where the branch ``$\pm $'' must be chosen for $\left\langle \mathcal{Q}_{1},%
\mathcal{Q}_{2}\right\rangle \gtrless 0$, respectively. In these formul\ae ,
$W\equiv W_{nBPS}$ is nothing but the Euclidean norm of the complex vector
of matter charges $D_{\widehat{i}}Z$ in local ``flat'' indices of $\mathbb{CP%
}^{n}$ \cite{ADOT-1,MS-FM-1}:
\begin{equation}
W=\sqrt{g^{i\overline{j}}D_{i}Z\overline{D}_{\overline{j}}\overline{Z}}=%
\sqrt{\sum_{\widehat{i}=1}^{n}\left| D_{\widehat{i}}Z\right| ^{2}},
\end{equation}
and it has been explicitly computed in \cite{Gnecchi-1}:
\begin{eqnarray}
W &=&\frac{1}{\sqrt{2}\left( 1-\left| t^{m}\right| ^{2}\right) }\left(
\delta ^{i\overline{j}}-t^{i}\overline{t}^{\overline{j}}\right) \cdot  \notag
\\
&&\cdot \left[ (q_{i}-ip^{i})\left( 1-\left| t^{m}\right| ^{2}\right)
+(q_{0}+ip^{0})\overline{t}^{\overline{i}}+(q_{r}-ip^{r})t^{r}\overline{t}^{%
\overline{i}}\right] \cdot  \notag \\
&&\cdot \left[ (q_{j}+ip^{j})\left( 1-\left| t^{m}\right| ^{2}\right)
+(q_{0}-ip^{0})t^{j}+(q_{n}+ip^{n})\overline{t}^{\overline{n}}t^{j}\right] ,
\label{W-non-BPS}
\end{eqnarray}
In the present paper, we are not going to deal with a general analysis of
Eqs. (\ref{MS-nBPS-n})-(\ref{W-non-BPS}), which will be given elsewhere.

\subsubsection{``\textit{Moduli Spaces''} of Multi-Center Flows}

We now briefly discuss the \textit{``moduli spaces''} of $p$-center non-BPS
solutions in \textit{minimally coupled} $\mathbb{CP}^{n}$ $\mathcal{N}=2$, $%
d=4$ models.

It is known \cite{Ferrara-Marrani-2} that for $p=1$ the \textit{``moduli
space''} is
\begin{equation}
\mathcal{M}_{nBPS,\mathbb{CP}^{n},p=1}=\frac{U(1,n-1)}{U(1)\times U(n-1)}\,.
\end{equation}
Its generalization to the case of $2\leqslant p\leqslant n$ centers is%
\footnote{%
We are grateful to R. Stora for an enlightening discussion on this point.}
\begin{equation}
\mathcal{M}_{nBPS,\mathbb{CP}^{n},p}=\frac{U(1,n-p)}{U(1)\times U(n-p)}\,.
\end{equation}
In order to prove this, we notice that the generic orbit of $p$ $(n+1)$%
-dimensional complex vectors $\left\{ \mathcal{X}_{\mathbf{a}}\right\} _{%
\mathbf{a}=1,...,p}$ in the $\mathbf{1+n}$ of $U(1,n)$ with $\mathbf{I}_{%
\mathbf{a}}=\mathcal{X}_{\mathbf{a}}\cdot \overline{\mathcal{X}}_{\mathbf{a}%
}<0$ $\forall \mathbf{a}$ (see \textit{e.g.} (\ref{r-1})) is
\begin{equation}
\mathcal{O}_{nBPS,\mathbb{CP}^{n},p}=\frac{U(1,n)}{U(1,n-p)}\,.
\end{equation}
With $p$ complex vectors $\left\{ \mathcal{X}_{\mathbf{a}}\right\} _{\mathbf{%
a}=1,...,p}$, one can build $p^{2}$ $U(1,n)$-invariants $\mathcal{X}_{%
\mathbf{a}}\cdot \overline{\mathcal{X}}_{\mathbf{b}}$ ($\mathbf{a},\mathbf{b}%
=1,...,p$; recall definition (\ref{deff})), corresponding to $p^{2}$ real
degrees of freedom. Thus, the following consistent counting holds:
\begin{equation}
p^{2}+\text{dim}_{\mathbb{R}}(\mathcal{O}_{nBPS,\mathbb{CP}%
^{n},p})=2p(n+1)\,,
\end{equation}
where $2p(n+1)$ is the number of real charge degrees of freedom pertaining
to $p$ $(n+1)$-dimensional vectors $\left\{ \mathcal{X}_{\mathbf{a}}\right\}
_{\mathbf{a}=1,...,p}$ of complexified charges (recall definition (\ref
{X-def})).

Therefore, ``flat directions'' (and thus \textit{``moduli spaces''}) for
non-BPS $p$-center flows in $\mathcal{N}=2$ $\mathbb{CP}^{n}$ models arise
only for $p<n$. In particular, for $p=2$ centers, one needs at least $n=3$.
Incidentally, this model is ``dual'' \cite{Gnecchi-1} to $\mathcal{N}=3$
supergravity with one matter multiplet (for a discussion of split flows and
marginal stability in extended $d=4$ supergravities, see \cite{MS-FM-1}).

\section{\label{BPS-t^3-Analysis}A Comparison :\newline
BPS MS \textit{and} AMS Walls in the $t^{3}$ Model}

We refer to the treatment of the BPS two-center solutions in $\mathcal{N}=2$%
, $d=4$ $t^{3}$ model, given in Sec. 5 of \cite{BD-1}. The symplectic charge
vectors $\mathcal{Q}_{1}$ and $\mathcal{Q}_{2}$ of the two centers are
chosen as follows ($u,q,v\in \mathbb{R}_{0}^{+}$):
\begin{eqnarray}
\mathcal{Q}_{1} &\equiv &\left( -p^{0},p^{1},q_{0},q_{1}/3\right) ^{T}\equiv
\left( v,0,0,q\right) \Rightarrow \mathcal{I}_{4}\left( \mathcal{Q}%
_{1}\right) >0;  \label{Q1} \\
\mathcal{Q}_{2} &\equiv &\left( -P^{0},P^{1},Q_{0},Q_{1}/3\right) ^{T}\equiv
\left( 0,0,u,0\right) \Rightarrow \mathcal{I}_{4}\left( \mathcal{Q}%
_{2}\right) =0,  \label{Q2}
\end{eqnarray}
yielding a mutual non-locality:
\begin{equation}
\left\langle \mathcal{Q}_{1},\mathcal{Q}_{2}\right\rangle =-uv<0.
\label{NonLocal}
\end{equation}
Thus, this is a case with ``large'' (and thus attractive) BPS center $1$,
and ``small'' ($1$-charge) center $2$.

Note that, as also observed in \cite{BD-1}, \textit{iff} $\mathcal{I}%
_{4}\left( \mathcal{Q}_{1}+\mathcal{Q}_{2}\right) >0$ the ``large'' BPS
single-center solution with charge vector $\mathcal{Q}=\mathcal{Q}_{1}+%
\mathcal{Q}_{2}$ would exist, as well. If this occurs, it should be pointed
out that
\begin{equation}
\sqrt{\mathcal{I}_{4}\left( \mathcal{Q}_{1}+\mathcal{Q}_{2}\right) }<\sqrt{%
\mathcal{I}_{4}\left( \mathcal{Q}_{1}\right) }+\sqrt{\mathcal{I}_{4}\left(
\mathcal{Q}_{2}\right) }=\sqrt{\mathcal{I}_{4}\left( \mathcal{Q}_{1}\right) }%
,  \label{Quart-in}
\end{equation}
namely that the two-center BPS solutions with charges $\mathcal{Q}_{1}$ and $%
\mathcal{Q}_{2}$, if it exists, has more entropy than the corresponding BPS
single-center solution with charge $\mathcal{Q}=\mathcal{Q}_{1}+\mathcal{Q}%
_{2}$. As discussed in Sec. \ref{Intro}, this is the opposite of what holds
for the BPS split flows in $\mathcal{N}=2$, $d=4$ \textit{minimally coupled}
models.

The corresponding holomorphic central charges read ($t\equiv b+ia$) \cite
{BD-1}
\begin{eqnarray}
Z\left( \mathcal{Q}_{1}\right) &\equiv &Z_{1}=3qt-vt^{3}=\left(
3q-vb^{2}+3a^{2}v\right) b+i\left( 3q-3b^{2}v+a^{2}v\right) a;  \label{Z1} \\
Z\left( \mathcal{Q}_{2}\right) &\equiv &Z_{2}=u.  \label{Z2}
\end{eqnarray}
Note that, within the conventions of \cite{BD-1}, the domain of definition
of the metric of the scalar manifold is $a\in \mathbb{R}_{0}^{+}$.

(\ref{Z1}) implies that
\begin{equation}
Z\left( \mathcal{Q}_{1}\right) =0\Leftrightarrow \left( b,a\right) =\left\{
\begin{array}{l}
\pm \left( \sqrt{3\frac{q}{v}},0\right) ; \\
\text{or} \\
\left( 0,0\right) ,
\end{array}
\right.
\end{equation}
which are both outside the domain of definition of the metric of the scalar
manifold itself. On the other hand, (\ref{Z2}) implies that $Z\left(
\mathcal{Q}_{2}\right) $ never vanishes (because $u>0$).

From (\ref{Z1}) and (\ref{Z2}), one can compute that
\begin{eqnarray}
\text{Re}\left( Z_{1}\overline{Z_{2}}\right) &=&\text{Re}\left( Z_{1}\right)
\text{Re}\left( Z_{2}\right) +\text{Im}\left( Z_{1}\right) \text{Im}\left(
Z_{2}\right) =\left[ 3q-v\left( b^{2}-3a^{2}\right) \right] ub;
\label{ReZ1Z2bar} \\
\text{Im}\left( Z_{1}\overline{Z_{2}}\right) &=&\text{Im}\left( Z_{1}\right)
\text{Re}\left( Z_{2}\right) -\text{Re}\left( Z_{1}\right) \text{Im}\left(
Z_{2}\right) =\left[ 3q-v\left( 3b^{2}-a^{2}\right) \right] ua.
\label{ImZ1Z2bar}
\end{eqnarray}
\medskip

Exploiting (\ref{ImZ1Z2bar}) one can compute
\begin{equation}
\text{Im}\left( Z_{1}\overline{Z_{2}}\right) =0\Leftrightarrow b^{2}=\frac{%
a^{2}}{3}+\frac{q}{v}\Leftrightarrow b=\pm \sqrt{\frac{a^{2}}{3}+\frac{q}{v}}%
,  \label{b^2}
\end{equation}
%%%%%%%%%%%%%%%%%%%%%%%%%%%%%%%%%%%%%%%%%%%%%%%%%%%%%%%%%%%%%%%%%%%%%
%%%%%%%%%%%%%%%%%%%%%%%%%%%%%%%%%%%%%%%%%%%%%%%%%%%%%%%%%%%%%%%%%%%%%
%%%%%%%%%%%%%%%%%%%%%%%%%%%%%%%%%%%%%%%%%%%%%%%%%%%%%%%%%%%%%%%%%%%%%
%%%%%%%%%%%%%%%%%%%%%%%%%%%%%%%%%%%%%%%%%%%%%%%%%%%%%%%%%%%%%%%%%%%%%
%%%%%%%%%%%%%%%%%%%%%%%%%%%%%%%%%%%%%%%%%%%%%%%%%%%%%%%%%%%%%%%%%%%%%
where the argument of the root is always positive. Notice that (\ref{b^2})
automatically implies $b^{2}>q/v$ which is a required condition to have a
well-defined axion $b$. Moreover, by combining Eqs. (\ref{b^2}) and (\ref
{ReZ1Z2bar}), one obtains
\begin{equation}
\text{Re}\left( Z_{1}\overline{Z_{2}}\right) =2ub\left( q+4a^{2}v\right) ,
\end{equation}
and thus it is immediate to realize that the sign of $b$ determines the very
nature of the wall itself. Indeed, by recalling the definitions (\ref{MS-BPS}%
) and (\ref{AMS-BPS}), the BPS MS and AMS walls $MS_{BPS}$ and $AMS_{BPS}$
can be computed to be given by:
\begin{eqnarray}
MS_{BPS}\left( \mathcal{Q}_{1},\mathcal{Q}_{2}\right) &:&b=\sqrt{\frac{a^{2}%
}{3}+\frac{q}{v}};  \label{BPS-MS-t^3-sol} \\
AMS_{BPS}\left( \mathcal{Q}_{1},\mathcal{Q}_{2}\right) &:&b=-\sqrt{\frac{%
a^{2}}{3}+\frac{q}{v}};  \label{BPS-AMS-t^3-sol}
\end{eqnarray}

%The definition of $AMS_{BPS}\left( \mathcal{Q}_{1},\mathcal{Q}_{2}\right) $
%given by (\ref{BPS-AMS-t^3}) is a combination of (\ref{ImZ1Z2bar=0}) and (%
%\ref{ReZ1Z2bar<0-I}) (\textit{or} (\ref{ReZ1Z2bar<0-II})). Due to (\ref
%{a-ineq}), it is clear that the solution $\mathbf{III}$ given by (\ref
%{ReZ1Z2bar<0-I}) is not consistent with (\ref{ImZ1Z2bar=0}). Thus, the
%solution defining $AMS_{BPS}\left( \mathcal{Q}_{1},\mathcal{Q}_{2}\right) $
%is given by the combination of (\ref{ImZ1Z2bar=0}) and solution $\mathbf{IV}$
%given by (\ref{ReZ1Z2bar<0-II}):
%\begin{equation}
%AMS_{BPS}\left( \mathcal{Q}_{1},\mathcal{Q}_{2}\right) :\left\{
%\begin{array}{l}
%a=\sqrt{3}\sqrt{b^{2}-\frac{q}{v}}; \\
%\\
%b<-\sqrt{3\frac{q}{v}}.
%\end{array}
%\right.   \label{BPS-AMS-t^3-sol}
%\end{equation}

Notice that by solving Im$\left( Z_{1}\overline{Z_{2}}\right) =0$ with
respect to the axion $b$ and plugging the solution into (\ref{ReZ1Z2bar}),
leads to the following expression for Re$\left( Z_{1}\overline{Z_{2}}\right)
$ at the points at which Im$\left( Z_{1}\overline{Z_{2}}\right) =0$ :
\begin{equation}
\left. \text{Re}\left( Z_{1}\overline{Z_{2}}\right) \right| _{\text{Im}%
\left( Z_{1}\overline{Z_{2}}\right) =0}=\left. \left[ 3q-v\left(
b^{2}-3a^{2}\right) \right] ub\right| _{a^{2}=3\left( b^{2}-\frac{q}{v}%
\right) }=8uvb\left( b^{2}-\frac{3}{4}\frac{q}{v}\right) ,
\end{equation}
which matches the expression given by Eq. (5.8) of \cite{BD-1}, but does not
have a definite sign.

\begin{figure}[h!]
\centering \epsfxsize= 11cm \epsfysize= 7cm %
\epsfbox{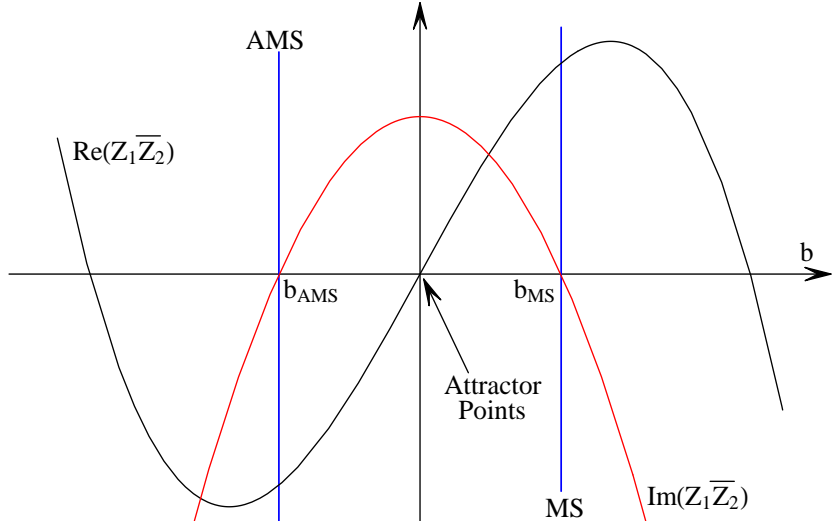} \caption{Plot of $\text{Im}\left(
Z_{1}\overline{Z_{2}}\right)$ (red curve) and $\text{Re}\left(
Z_{1}\overline{Z_{2}}\right)$ (black curve) as a
function of $b$. MS and AMS walls (blue lines) are identified by $b_{MS}=%
\protect\sqrt{a^2/3+q/v}$ and $b_{AMS}=-\protect\sqrt{a^2/3+q/v}$.
It is manifest that the (physically sensible) flow connecting MS to
AMS wall must cross the instability region.}
\end{figure}

In general, the flow is directed from stability to instability. Assuming the
flowing dynamics from the BPS MS wall $MS_{BPS}\left( \mathcal{Q}_{1},%
\mathcal{Q}_{2}\right) $ towards the BPS AMS wall $AMS_{BPS}\left( \mathcal{Q%
}_{1},\mathcal{Q}_{2}\right) $
\begin{equation}
MS_{BPS}\left( \mathcal{Q}_{1},\mathcal{Q}_{2}\right) :\left\{
\begin{array}{l}
\text{Im}\left( Z_{1}\overline{Z_{2}}\right) =0; \\
\\
\text{Re}\left( Z_{1}\overline{Z_{2}}\right) >0.
\end{array}
\right. \rightsquigarrow AMS_{BPS}\left( \mathcal{Q}_{1},\mathcal{Q}%
_{2}\right) :\left\{
\begin{array}{l}
\text{Im}\left( Z_{1}\overline{Z_{2}}\right) =0; \\
\\
\text{Re}\left( Z_{1}\overline{Z_{2}}\right) <0.
\end{array}
\right.
\end{equation}
to be continuous, then surely the flow itself will crash into a point in
which Re$\left( Z_{1}\overline{Z_{2}}\right) =0$. The locations at which
this occurs can be identified by
\begin{gather}
\text{Re}\left( Z_{1}\overline{Z_{2}}\right) =0\Leftrightarrow \left[
3q-v\left( b^{2}-3a^{2}\right) \right] ub=0\overset{u\in \mathbb{R}_{0}^{+}}{%
\Leftrightarrow }\left[ 3q-v\left( b^{2}-3a^{2}\right) \right] b=0  \notag \\
\Updownarrow  \notag \\
\left\{
\begin{array}{l}
\mathbf{i})~b=0; \\
\text{\textit{or}} \\
\mathbf{ii})~b^{2}=3\left( a^{2}+\frac{q}{v}\right) \Leftrightarrow b=\pm
\sqrt{3}\sqrt{a^{2}+\frac{q}{v}}.
\end{array}
\right.  \label{ReZ1Z2bar=0}
\end{gather}

In order to understand if the flow connecting $MS_{BPS}\left( \mathcal{Q}%
_{1},\mathcal{Q}_{2}\right) $ to $AMS_{BPS}\left( \mathcal{Q}_{1},\mathcal{Q}%
_{2}\right) $ belongs to the BPS stability region
\begin{equation}
\mathcal{S}_{BPS}\left( \mathcal{Q}_{1},\mathcal{Q}_{2}\right) :\left\langle
Q_{1},Q_{2}\right\rangle \text{Im}\left( Z_{1}\overline{Z_{2}}\right) >0\,,
\label{StabCond}
\end{equation}
one has to check if the condition
\begin{equation}
\text{Im}\left( Z_{1}\overline{Z_{2}}\right) =\left[ 3q-v\left(
3b^{2}-a^{2}\right) \right] ua<0  \label{RedStabCond}
\end{equation}
holds, since (\ref{StabCond}) reduces to (\ref{StabCond}) by using (\ref
{NonLocal}). By plugging the solutions $\mathbf{i}$ and $\mathbf{ii}$ of (%
\ref{ReZ1Z2bar=0}) into (\ref{ImZ1Z2bar}), one finds:
\begin{eqnarray}
\text{at~solution~}\mathbf{i~}\left( \notin \mathcal{S}_{BPS}\left( \mathcal{%
Q}_{1},\mathcal{Q}_{2}\right) \right) &:&\left. \text{Im}\left( Z_{1}%
\overline{Z_{2}}\right) \right| _{\mathbf{i}}=\left( 3q+va^{2}\right)
ua>0\,\rightarrow \text{unstable~};  \notag \\
&& \\
\text{at~solution~}\mathbf{ii~}\left( \in \mathcal{S}_{BPS}\left( \mathcal{Q}%
_{1},\mathcal{Q}_{2}\right) \right) &:&\left. \text{Im}\left( Z_{1}\overline{%
Z_{2}}\right) \right| _{\mathbf{ii}}=-2\left( 4va^{2}+3q\right)
ua<0\,\rightarrow \text{stable~}.  \notag \\
&&
\end{eqnarray}
On top of that, the last equations clearly prove that
\begin{equation}
\text{Re}\left( Z_{1}\overline{Z_{2}}\right) =0\nRightarrow \text{Im}\left(
Z_{1}\overline{Z_{2}}\right) =0,  \label{not-impl-1}
\end{equation}
and thus Re$\left( Z_{1}\overline{Z_{2}}\right) =0$ does \textit{not} imply $%
Z\left( \mathcal{Q}_{1}\right) =0$ nor $Z\left( \mathcal{Q}_{2}\right) =0$.
One can reach the same conclusion by recalling (\ref{Z1}), from which we
gain (from (\ref{Z2}), $Z_{2}=u\in \mathbb{R}_{0}^{+}$)
\begin{eqnarray}
\text{at~solution~}\mathbf{i~}\left( \notin \mathcal{S}_{BPS}\left( \mathcal{%
Q}_{1},\mathcal{Q}_{2}\right) \right) &:&\left. Z_{1}\right| _{\mathbf{i}%
}=i\left( 3q+va^{2}\right) a\neq 0; \\
\text{at~solution~}\mathbf{ii~}\left( \in \mathcal{S}_{BPS}\left( \mathcal{Q}%
_{1},\mathcal{Q}_{2}\right) \right) &:&\left. Z_{1}\right| _{\mathbf{ii}%
}=-i\left( 8va^{2}+6q\right) a\neq 0.
\end{eqnarray}

Thus, the flow encounters points at which Re$\left( Z_{1}\overline{Z_{2}}%
\right) =0$, but at which Im$\left( Z_{1}\overline{Z_{2}}\right) \neq 0$,
and also both $Z_{1}$ and $Z_{2}$ are \textit{non-vanishing}.

Note that, in order to go from the BPS MS wall $MS_{BPS}\left( \mathcal{Q}%
_{1},\mathcal{Q}_{2}\right) $ to the BPS AMS wall \linebreak $%
AMS_{BPS}\left( \mathcal{Q}_{1},\mathcal{Q}_{2}\right) $, the flow
necessarily cross the \textit{instability} region $-\sqrt{\frac{q}{v}}%
\leqslant b\leqslant \sqrt{\frac{q}{v}}$. In particular, the flow crosses
the axis $a$ (at which $b=0$, and thus Re$\left( Z_{1}\overline{Z_{2}}%
\right) =0$; see solution $\mathbf{i}$ of (\ref{ReZ1Z2bar=0})).

The situation is depicted in Fig. 2.

\section{\label{Conclusion}Conclusion}

The analysis carried out for \textit{minimally coupled} Maxwell-Einstein
supergravity is rather different from the one holding for $\mathcal{N}=2$
special K\"{a}hler geometries based on cubic prepotential \cite{David}, even
though it exhibits many general properties of the split attractor flow for
multi-center BHs.
%The analysis carried out for \textit{minimally coupled} Maxwell-Einstein
%supergravity, although it exhibits many properties of the split attractor
%flow for multi-center BHs, it is rather different from the one holding for $%
%\mathcal{N}=2$ special K\"{a}hler geometries based on cubic prepotential
%\cite{David}.
The properties of the latter have been considered to a large extent in the
literature, as they are related to Calabi-Yau compactifications.

On the other hand, $\mathcal{N}=3$ \cite{N=3} supergravity is expected to
have a split flow analysis analogous to the one studied for \textit{%
minimally coupled} $\mathcal{N}=2$ models in the present paper. Indeed, such
a theory also has a duality quadratic invariant $\mathcal{I}_{2}$, with the
charges sitting in the fundamental representation of the duality group $%
U(3,n)$ \cite{ADF-U-duality-d=4,Gnecchi-1}. Furthermore, the $\mathcal{N}=3$
$1$-modulus supergravity is ``dual'' to the \textit{minimally coupled} $3$%
-moduli ($\mathbb{CP}^{3}$) $\mathcal{N}=2$ theory, with the BPS and non-BPS
supersymmetry features interchanged \cite{Gnecchi-1}.

For theories with a duality invariant $\mathcal{I}_{4}$ which is quartic in
the charges, the analysis is more involved, because the charge orbits have a
more intricate structure. These latter theories are expected to exhibit
various phenomena, such as ``recombination walls'' \cite{ADMJ-2} and
``entropy enigmas'' \cite{DM-1,DM-2,David}, which are not present in the
class of theories analyzed in this work.

It is worth of notice that (non-compact forms of) $\mathbb{CP}^{n}$ spaces
as moduli spaces of string compactifications have appeared in the
literature, either as particular subspaces of complex structure deformations
of certain Calabi-Yau manifold \cite{CDR,Dixon:1989fj} or as moduli spaces
of some asymmetric orbifolds of Type II superstrings \cite{FK}--\nocite
{FF,DH}\cite{KK}, or of orientifolds \cite{Frey}.

\section*{Acknowledgments}

S.F. would like to thank Frederik Denef for very useful correspondence. The
work of S. F. is supported by the ERC Advanced Grant no. 226455, \textit{%
``Supersymmetry, Quantum Gravity and Gauge Fields''} (\textit{SUPERFIELDS})
and in part by DOE Grant DE-FG03-91ER40662.

E. O. would like to thank the Theory Division at CERN and the ERC Advanced
Grant no. 226455, \textit{``Supersymmetry, Quantum Gravity and Gauge Fields''%
} (\textit{SUPERFIELDS}) for financial support and kind hospitality.

The work of A. M. has been supported by an INFN visiting Theoretical
Fellowship at SITP, Stanford University, CA, USA.

%\newpage
\appendix \setcounter{equation}0

\section{\label{App-I}$\mathbf{I}_{s}>0$ as Condition of Existence for the
BPS MS Wall\newline
in $\mathbb{CP}^{n}$ Models (Cases \textbf{1} and \textbf{2})}

In this Appendix, we detail the derivation of the results (\ref{q0Q0>0})-(%
\ref{q0Q0<0}), relating the sign of the $U\left( 1,1\right) $-invariant $%
\mathbf{I}_{s}$ to the existence of the BPS MS wall \textit{or} of the BPS
AMS wall.

Through the conditions (\ref{Q2-BPS-large-cond}) and (\ref{impl}) and the
definition (\ref{def-a_BPS}), the general solution to the condition of
compatibility of the metric constraint (\ref{metric-domain}) with the
condition Im$\left( Z_{1}\overline{Z_{2}}\right) =0$:
\begin{equation}
\left\{
\begin{array}{l}
a^{2}+b^{2}<1; \\
\\
\text{Im}\left( Z_{1}\overline{Z_{2}}\right) =0
\end{array}
\right. \Leftrightarrow a^{2}+\frac{\mathbf{I}_{1}\mathbf{I}_{2}-\mathbf{I}%
_{s}^{2}}{\left( \mathbf{I}_{s}^{2}+\mathbf{I}_{a}^{2}-\mathbf{I}_{1}\mathbf{%
I}_{2}\right) }<0  \label{y-ineq}
\end{equation}
reads (see also (\ref{def-A-call}))
\begin{equation}
-a_{BPS}<a<a_{BPS}.  \label{BPS-y-sol}
\end{equation}
Therefore, due to (\ref{Q2-BPS-large-cond}), the following ordering on the $%
a $-axis holds:
\begin{eqnarray}
-\left| \frac{Q_{0}}{P^{1}}\right| &<&-a_{BPS}<a<a_{BPS}<\left| \frac{Q_{0}}{%
P^{1}}\right| ;  \label{BPS-y-order} \\
\left| \frac{Q_{0}}{P^{1}}\right| &=&\frac{\left| \mathbf{I}_{s}\right| }{%
\sqrt{\mathbf{I}_{s}^{2}+\mathbf{I}_{a}^{2}-\mathbf{I}_{1}\mathbf{I}_{2}}}.
\label{BPS-y-order-2}
\end{eqnarray}

After defining $\mathcal{A}$ through (\ref{def-A-call}), let us now analyse
all sign possibilities for the relevant quantities
\begin{eqnarray}
Q_{0}P^{1} &=&\pm \left| \frac{\mathbf{I}_{s}}{\mathbf{I}_{1}}\right| \sqrt{%
\mathbf{I}_{s}^{2}+\mathbf{I}_{a}^{2}-\mathbf{I}_{1}\mathbf{I}_{2}};
\label{PA-3} \\
q_{0}P^{1} &=&\pm \sqrt{\mathbf{I}_{s}^{2}+\mathbf{I}_{a}^{2}-\mathbf{I}_{1}%
\mathbf{I}_{2}},  \label{PA-4}
\end{eqnarray}
where the ``$\pm $'' branches in (\ref{PA-3}) and (\ref{PA-4}) are clearly
independent.

\begin{itemize}
\item  Let us start by choosing the branch ``$+$'' in (\ref{PA-3}). If one
chooses the branch ``$+$'' also in (\ref{PA-4}), then the exploitation of (%
\ref{MS-BPS})-(\ref{AMS-BPS}) yields
\begin{equation}
\begin{array}{l}
MS_{BPS}=\mathcal{A}; \\
\nexists AMS_{BPS}.
\end{array}
\label{PPA-1}
\end{equation}
On the other hand, if one chooses the branch ``$-$'' in (\ref{PA-4}), then (%
\ref{MS-BPS})-(\ref{AMS-BPS}) imply
\begin{equation}
\begin{array}{l}
\nexists MS_{BPS}; \\
AMS_{BPS}=\mathcal{A}.
\end{array}
\label{PPA-2}
\end{equation}

\item  Let us now consider the branch ``$-$'' in (\ref{PA-3}). If one
chooses the branch ``$-$'' in (\ref{PA-4}), then (\ref{MS-BPS})-(\ref
{AMS-BPS}) imply (\ref{PPA-2}). On the other hand, if the branch ``$-$'' is
chosen also in (\ref{PA-4}), (\ref{MS-BPS})-(\ref{AMS-BPS}) yield to the
result (\ref{PPA-1}).
\end{itemize}

By summarizing the various results, it is immediate to realize that only the
sign of $q_{0}Q_{0}=\mathbf{I}_{s}$ (recall (\ref{n=1-BPS-invs})) is
relevant: when this quantity is positive, the BPS MS wall exists, but
\textit{not} the AMS wall, and \textit{vice versa} when such a quantity is
negative, as given by Eqs. (\ref{q0Q0>0})-(\ref{q0Q0<0}).

\textit{Mutatis mutandis}, an analogous treatment holds for the case $%
n\geqslant 2$, leading to the results (\ref{q0Q0>0-n})-(\ref{q0Q0<0-n}).

\setcounter{equation}0

\section{\label{App-II}$\mathbf{I}_{s}<0$ as Condition of Existence for the
non-BPS MS Wall\newline
in the $\mathbb{CP}^{1}$ Model}

In this Appendix, we detail the derivation of the results (\ref{q1Q1>0})-(%
\ref{q1Q1<0}), relating the sign of the $U\left( 1,1\right) $-invariant $%
\mathbf{I}_{s}$ to the existence of the non-BPS MS wall \textit{or} of the
BPS AMS wall. Essentially, all the treatment of this Appendix can be
obtained from the treatment given in App. \ref{App-I} (for $n=1$) by
applying the transformation (\ref{Q-transf-++++}) to both $\mathcal{Q}_{1}$
and $\mathcal{Q}_{2}$.

Through the conditions (\ref{Q2-nBPS-cond}) and (\ref{impl-nBPS}) and the
definition (\ref{def-a_nBPS}), the general solution to the condition of
compatibility of the metric constraint (\ref{metric-domain}) with the
condition Im$\left( D_{\widehat{t}}Z_{1}\overline{D_{\widehat{t}}Z_{2}}%
\right) =0$
\begin{equation}
\left\{
\begin{array}{l}
a^{2}+b^{2}<1; \\
\\
\text{Im}\left( D_{\widehat{t}}Z_{1}\overline{D_{\widehat{t}}Z_{2}}\right) =0
\end{array}
\right. \Leftrightarrow a^{2}+\frac{\left( \mathbf{I}_{s}^{2}-\mathbf{I}_{1}%
\mathbf{I}_{2}\right) }{\left( \mathbf{I}_{1}\mathbf{I}_{2}-\mathbf{I}%
_{s}^{2}-\mathbf{I}_{a}^{2}\right) }<0  \label{y-ineq-nBPS}
\end{equation}
reads (see also (\ref{def-a_nBPS}))
\begin{equation}
-a_{nBPS}<a<a_{nBPS}.  \label{nBPS-y-sol}
\end{equation}
Therefore, due to (\ref{Q2-nBPS-cond}), the following ordering on the $a$%
-axis holds:
\begin{eqnarray}
-\left| \frac{Q_{1}}{P^{0}}\right| &<&-a_{nBPS}<a<a_{nBPS}<\left| \frac{Q_{1}%
}{P^{0}}\right| ;  \label{nBPS-y-order} \\
\left| \frac{Q_{1}}{P^{0}}\right| &=&\frac{\left| \mathbf{I}_{s}\right| }{%
\sqrt{\mathbf{I}_{s}^{2}+\mathbf{I}_{a}^{2}-\mathbf{I}_{1}\mathbf{I}_{2}}}.
\label{nBPS-y-order-2}
\end{eqnarray}

After defining $\mathcal{B}$ through (\ref{B-call}), let us now analyse all
sign possibilities for the relevant quantities
\begin{eqnarray}
Q_{1}P^{0} &=&\pm \left| \frac{\mathbf{I}_{s}}{\mathbf{I}_{1}}\right| \sqrt{%
\mathbf{I}_{s}^{2}+\mathbf{I}_{a}^{2}-\mathbf{I}_{1}\mathbf{I}_{2}};
\label{PA-5} \\
q_{1}P^{0} &=&\pm \sqrt{\mathbf{I}_{s}^{2}+\mathbf{I}_{a}^{2}-\mathbf{I}_{1}%
\mathbf{I}_{2}},  \label{PA-6}
\end{eqnarray}
where the ``$\pm $'' branches in (\ref{PA-5}) and (\ref{PA-6}) are clearly
independent.

\begin{itemize}
\item  Let us start by choosing the branch ``$+$'' in (\ref{PA-5}). If one
chooses the branch ``$+$'' also in (\ref{PA-6}), then the exploitation of (%
\ref{MS-nBPS})-(\ref{AMS-nBPS}) yields
\begin{equation}
\begin{array}{l}
MS_{nBPS}=\mathcal{B}; \\
\nexists AMS_{nBPS}.
\end{array}
\label{PPA-3}
\end{equation}
On the other hand, if one chooses the branch ``$-$'' in (\ref{PA-6}), then (%
\ref{MS-nBPS})-(\ref{AMS-nBPS}) imply
\begin{equation}
\begin{array}{l}
\nexists MS_{nBPS}; \\
AMS_{nBPS}=\mathcal{B}.
\end{array}
\label{PPA-4}
\end{equation}

\item  Let us now consider the branch ``$-$'' in (\ref{PA-5}). If one
chooses the branch ``$-$'' in (\ref{PA-6}), then (\ref{MS-nBPS})-(\ref
{AMS-nBPS}) imply (\ref{PPA-4}). On the other hand, if the branch ``$-$'' is
chosen also in (\ref{PA-6}), (\ref{MS-nBPS})-(\ref{AMS-nBPS}) yield to the
result (\ref{PPA-3}).
\end{itemize}

By summarizing the various results, it is immediate to realize that only the
sign of $q_{1}Q_{1}=-\mathbf{I}_{s}$ (recall (\ref{n=1-nBPS-invs})) is
relevant: when $\mathbf{I}_{s}<0$, the non-BPS MS wall exists, but \textit{%
not} the AMS wall, and \textit{vice versa} when $\mathbf{I}_{s}>0$, as given
by Eqs. (\ref{q1Q1>0})-(\ref{q1Q1<0}).

\end{document}